\documentclass[aps, nofootinbib, prd, superscriptaddress, tightenlines,
preprintnumbers, notitlepage, twocolumn, floatfix]{revtex4-1}
\usepackage{tabularx}
\usepackage{graphicx}
\usepackage{amssymb,bm,tensor}
\usepackage{textcomp}
\usepackage{xcolor}
\usepackage{amsmath}
\usepackage[varg]{txfonts}
\usepackage{enumerate}
\usepackage{mathtools}
\usepackage[normalem]{ulem}
\usepackage{bm} 
\usepackage[OT1]{fontenc}
\usepackage[colorlinks]{hyperref}

\newcommand{\ReplyA}[1]{#1} 
\newcommand{\ReplyB}[1]{#1} 

\newcommand{\KIAA}{\affiliation{Kavli Institute for Astronomy and
Astrophysics, Peking University, Beijing 100871, China}}
\newcommand{\MPIfR}{\affiliation{Max-Planck-Institut f\"ur Radioastronomie,
Auf dem H\"ugel 69, D-53121 Bonn, Germany}}
\newcommand{\USTC}{\affiliation{Interdisciplinary Center for Theoretical Study, University of Science and Technology of China, Hefei, Anhui 230026, China and Peng Huanwu Center for Fundamental Theory, Hefei, Anhui 230026, China}}
\newcommand{\NAOC}{\affiliation{National Astronomical Observatories,
Chinese Academy of Sciences, Beijing 100012, China}}

\begin{document}

\title{New Graviton Mass Bound from Binary Pulsars}
\date{\today}
\author{Lijing Shao}\email{lshao@pku.edu.cn}\KIAA\MPIfR\NAOC
\author{Norbert Wex}\MPIfR
\author{Shuang-Yong Zhou}\USTC
\preprint{USTC-ICTS/PCFT-20-15}

\begin{abstract} 
  In Einstein's general relativity, gravity is mediated by a massless
  metric field. The extension of general relativity to consistently include
  a mass for the graviton has profound implications for gravitation and
  cosmology. Salient features of various massive gravity theories can be
  captured by Galileon models, the simplest of which is the cubic Galileon.
  The presence of the Galileon field leads to additional gravitational
  radiation in binary pulsars where the Vainshtein mechanism is less
  suppressed than its fifth-force counterpart, which deserves a detailed
  confrontation with observations. We prudently choose fourteen well-timed
  binary pulsars, and from their intrinsic orbital decay rates we put a new
  bound on the graviton mass, \ReplyB{$m_g \lesssim 2 \times 10^{-28}\,{\rm
  eV}/c^2$ at the 95\% confidence level}, assuming a flat prior on $\ln
  m_g$. It is equivalent to a bound on \ReplyB{the graviton Compton wavelength
  $\lambda_g \gtrsim 7 \times 10^{21}$\,m.} Furthermore, we extensively
  simulate times of arrival for pulsars in orbit around stellar-mass black
  holes and the supermassive black hole at the Galactic center, and
  investigate their prospects in probing the cubic Galileon theory in the
  near future.
\end{abstract}

\maketitle

\section{Introduction}
\label{sec:intro}

The late-time accelerated cosmic expansion poses a profound challenge for
modern physics, which is known as the dark energy problem~\cite{Li:2011sd,
Gleyzes:2013ooa, Joyce:2014kja}. From observations, we know that dark
energy manifests at lengthscales significantly larger than the Galactic
size, or in field-theoretic terminology, in the {\it infrared} regime. Dark
energy is often hypothesized as a cosmological constant in the standard
$\Lambda$CDM model~\cite{Weinberg:1988cp}, but its real nature remains
elusive. Astrophysical objects (in particular, the type Ia supernovae) in
the relatively nearby Universe and the cosmic microwave background in the
early Universe provide two classes of independent probes to measure the
Hubble expansion parameter at today, $H_0$. Recent observations from them,
however, have inferred inconsistent values of $H_0$ at a significance level
of 4.4\,$\sigma$~\cite{Aghanim:2018eyx, Birrer:2018vtm, Riess:2019cxk}. The
discrepancy aggravates the dark energy puzzle, and in the meantime has
triggered tremendous interest in searching for new physics beyond the
standard paradigm.

One of the main approaches to explain the dark energy phenomena involves
infrared modifications to the canonical gravity theory, general relativity
(GR)~\cite{Clifton:2011jh, Berti:2015itd, Heisenberg:2018vsk}.
Modifications usually introduce extra field contents, with a scalar degree
of freedom being the simplest and the most widely investigated in
literature. However, such a new scalar is likely to bring in a {\it fifth
force}~\cite{Clifton:2011jh, Berti:2015itd, Heisenberg:2018vsk}, which is
stringently constrained by observations in the Solar
system~\cite{Will:2014kxa} and binary pulsars~\cite{Wex:2014nva,
Shao:2016ezh, Yunes:2010qb, Yagi:2013qpa, Yagi:2013ava, Shao:2018klg}.
Therefore, to successfully account for the accelerated expansion of the
Universe, we need a modified gravity theory where the theory gives rise to
order-one corrections at cosmological scales but deviations from GR are
extremely suppressed in the Solar system, which makes nonlinearity a
crucial ingredient in the theory. For a class of such infrared
modifications of gravity, including the Galileon models~\cite{Luty:2003vm,
Nicolis:2008in}, this is achieved by the Vainshtein
mechanism~\cite{Vainshtein:1972sx, Babichev:2013usa}, by which the new
scalar becomes nonlinearly coupled in the local dense environment, thus
suppressing the fifth force in the Solar system. The length scale within
which the scalar becomes strongly-coupled is called the Vainshtein radius,
$r_\star$, and it is only outside of the Vainshtein radius that the linear
perturbations can be trusted. It is important that when dealing with models
with the Vainshtein mechanism the full nonlinear theory, as opposed to the
linear theory, needs to be solved to make correct physical interpretations.

The Vainshtein mechanism is intimately related to massive gravity, and the
accelerated cosmic expansion may be due to a condensate of gravitons with a
Hubble-scale mass~\cite{deRham:2014zqa, deRham:2016nuf}. It was pointed out
in 1970s that the unique (Lorentz invariant) linear theory of massive
gravity deviates from GR by order-one corrections, known as the van
Dam-Veltman-Zakharov discontinuity~\cite{vanDam:1970vg, Zakharov:1970cc,
Iwasaki:1971uz}. Vainshtein soon after suggested that this can not be used
to rule out massive gravity and rather one needs to solve the nonlinear
theory in environments such as the Solar system to get the right
prediction~\cite{Vainshtein:1972sx}. In other words, while the conventional
helicity-2 modes of GR become strongly coupled at the Schwarzschild radius,
the extra modes of massive gravity becomes strongly coupled within a much
larger Vainshtein radius for the same central mass. \ReplyB{To extract the
most strongly coupled extra modes in massive gravity, one takes the
decoupling limit as per the de Rham-Gabadadze-Tolley (dRGT) tuning
\cite{deRham:2010kj, deRham:2010ik}},
\begin{equation}
  \label{eq:decouple:limit} 
  m_g \to 0 \,, \quad M_{\rm Pl}\to \infty \,, \quad
  \Lambda=\left(m_g^2M_{\rm Pl}\right)^{1/3} \to ~ {\rm fixed} \,,
\end{equation}
where $m_g$ is the graviton mass and $M_{\rm Pl} \equiv 1/\sqrt{8\pi G} $
is the reduced Planck mass, and obtains a scalar effective field theory
with the Galileon symmetry,
\begin{equation} \label{eq:Galileon}
  \pi_s \to \pi_s + a + b_\mu x^\mu \,,
\end{equation}
where $\pi_s$ is the Galileon field and $a$ and $b_\mu$ are constants. The
decoupling limit (\ref{eq:decouple:limit}) scales away the small effects
due to other modes and let us focus on the physics most important near the
scale of $\Lambda$. In the presence of matter sources, we also take the
limit where the energy momentum tensor $T_{\mu\nu}$ goes to infinity but
the ratio between $T_{\mu\nu}/M_{\rm Pl}$ fixed. The Galileon scalar then
couples to the trace of $T_{\mu\nu}$ and encapsulates the salient features,
including the Vainshtein mechanism, of the extra modes in massive
gravity~\cite{Luty:2003vm, Nicolis:2008in}.

In this paper, we study the simplest model that exhibits the Vainshtein
mechanism, namely the cubic Galileon~\cite{Luty:2003vm}. This model is the
decoupling limit of the Dvali-Gabadadze-Porrati braneworld
model~\cite{Dvali:2000rv, Luty:2003vm}, where the graviton acquires a
so-called {\it soft} mass from embedding the 3-brane Universe in a 4-D bulk
with an Einstein-Hilbert term. The Galileon models are also the decoupling
limit of the recently discovered dRGT
model~\cite{deRham:2010kj, deRham:2010ik}, a unique nonlinear (Lorentz
invariant) massive gravity with a so-called {\it hard} mass. The
bi-gravity~\cite{Hassan:2011zd} or multi-gravity extension of the dRGT
model also leads to a bi-Galileon or multi-Galileon
theory~\cite{Padilla:2010de, Padilla:2010tj, Padilla:2010ir}.
\ReplyB{Therefore, the cubic Galileon model is often taken as a {\it proxy}
to cover all the Lorentz invariant massive gravity models, though by no
means it encodes all aspects of a complete theory of massive gravity where
other terms, for example the quartic Galileon term, might appear.}

Horndeski theory~\cite{Horndeski:1974wa}, the generalized scalar-tensor
theory with up to second derivatives in field equations, can be re-derived
in the Galileon framework~\cite{Deffayet:2011gz}. It is worth to mention
that, while a large class of Horndeski models have been ruled out by the
coincident observation of the gravitational-wave
signal~\cite{TheLIGOScientific:2017qsa} and the electromagnetic
counterpart~\cite{Monitor:2017mdv} from a binary neutron star inspiral
GW170817~\cite{Baker:2017hug, Ezquiaga:2017ekz,
Creminelli:2017sry},\footnote{Care, however, must be taken to interpret
this result, as the observed gravitational-wave frequencies are very close
to the cutoff of the Horndeski theory as an effective field
theory~\cite{deRham:2018red}.} the cubic Galileon subset of Horndeski
theory, though being a simple strawman model, is still comfortably alive.
Consequently, it is intriguing to study the cubic Galileon in the light of
recent research activities in the field.

The fifth force effects of massive gravity are often screened by the
Vainshtein mechanism so effectively that the existing constraints in the
dense environment can be easily evaded~\cite{Dvali:2002vf}, and only at a
cosmological density in the infrared regime can the theory deviate
significantly from GR, accounting for the dark energy. \ReplyB{However,
\citet{deRham:2012fw} has found that, in binary pulsar systems, the
suppression factor in the extra gravitational radiation due to the Galileon
mode is {\it less} than the suppression factor in the static fifth-force
effect (see Sec.~\ref{sec:theory}).} Therefore, it becomes extremely
interesting to check with the existing tests related to the gravitational
radiation in binary pulsars for the cubic Galileon model.

In this work, we present a thorough phenomenological study of the Galileon
radiation for binary pulsar systems. We carefully choose fourteen
well-timed binary pulsars to put constraints on the theory parameter of the
cubic Galileon. Among these pulsars, recent observations of the double
pulsar PSR~J0737$-$3039A~\cite{Kramer:2006nb, Kramer:2016kwa} give \ReplyB{the
strongest bound on the graviton mass, $m_g \lesssim 3 \times
10^{-28}\,{\rm eV}/c^2$ at the 95\% confidence level (C.L.).} Furthermore, a
combination of all fourteen pulsars in the Bayesian framework gives,
\begin{equation}
  \ReplyB{ m_g \lesssim 2 \times 10^{-28}\,{\rm eV}/c^2 \quad \mbox{(95\% C.L.)} } \,,
\end{equation}
with a flat prior on $\ln m_g$. It translates into a limit on \ReplyB{the
graviton Compton wavelength $\lambda_g \gtrsim 7 \times 10^{21}$\,m.}

The paper is organized as follows. In the next section, we review the
basics for the Galileon radiation~\cite{deRham:2012fw}. In
Sec.~\ref{sec:psr}, systematic studies are carried out to understand the
dependence of the Galileon radiation on system parameters and the figure of
merit to test it. Based on these studies, we choose fourteen binary pulsars
to cast tight constraints on the graviton mass. Limits are obtained from
individual pulsars as well as a combination of them. Moreover in
Sec.~\ref{sec:bhs}, with a set of simulated times of arrival for
near-future radio telescopes, we investigate the prospects in using pulsars
around a stellar-mass black hole (BH) companion~\cite{Wex:1998wt,
Liu:2014uka, Seymour:2018bce} and the supermassive BH at the Galactic
center (namely, the Sgr\,A$^*$)~\cite{Liu:2011ae, Psaltis:2015uza,
Bower:2018mta, Bower:2019ads} to constrain the cubic Galileon theory. The
last section presents some discussion and briefly summarizes the paper.

Throughout the paper, we implicitly assume units where $\hbar = c = 1$,
except for a couple of places where $\hbar$ and $c$ are restored for
readers' convenience.

\section{Theory}\label{sec:theory}

Since the Hulse-Taylor pulsar provided the first indirect evidence for the
existence of gravitational waves~\cite{Taylor:1979zz}, many more binary
pulsars have been playing an important role in probing the property of the
gravitational radiation in alternative gravity
theories~\cite{Freire:2012mg, Wex:2014nva, Yagi:2015oca, Kramer:2016kwa,
Shao:2017gwu, Zhao:2019suc}. We briefly review the gravitational radiation
for a binary pulsar in GR and in cubic Galileon theory in Sec.~\ref{sec:GR}
and Sec.~\ref{sec:Galileon:theory} respectively.

\subsection{General Relativity} \label{sec:GR}

Consider a binary system with component masses $m_1$ and $m_2$ in an orbit
with a semimajor axis $a$ and an eccentricity $e$. Due to the finite
propagating velocity of gravity, at the leading order the binary loses
energy by radiating off gravitational waves with an emitting
power~\cite{Peters:1963ux},
\begin{equation}
    \label{eq:power:GR}
    {\cal P}_{\rm GR}=\frac{32\eta^2}{5c^5} \frac{ G^4 M^5 }{a^{5}}
    \left(1+\frac{73}{24} e^{2}+\frac{37}{96} e^{4}\right) \left( 1-e^2
    \right)^{-7/2} \,,
\end{equation}
where the total mass $M \equiv m_1 + m_2$; the symmetric mass ratio $\eta
\equiv m_1 m_2 / M^2$; $G$ and $c$ are the gravitational constant and the
speed of light respectively.

From the Kepler's third law for a binary system, we have
\begin{equation}\label{eq:Kepler}
  n_b^2 a^3 = G M \,,
\end{equation}
where $n_b \equiv 2\pi/P_b$ with $P_b$ the orbital period. The
nonrelativistic orbital energy at the Newtonian order for the binary reads,
\begin{equation}\label{eq:orbit:energy}
  E_{b} = - \frac{\eta GM^2}{2a} \,.
\end{equation}
Taking time derivatives in Eqs.~(\ref{eq:Kepler}) and
(\ref{eq:orbit:energy}), we have
\begin{equation}
    \frac{\dot a}{a} = \frac{2}{3} \frac{\dot P_b}{P_b} \,,
\end{equation}
and
\begin{equation}
  \frac{\dot E_{b}}{E_{b}} = - \frac{\dot a}{a} \,.
\end{equation}
Finally, using the energy conservation law in GR, ${\cal P}_{\rm GR} =-\dot
E_{b} $, we have~\cite{Peters:1963ux},
\begin{equation}\label{eq:Pbdot:GR}
  \dot{P}_{b}^{\rm GR}=-\frac{192 \pi }{5c^5} \eta (GM)^{5/3} n_b^{5 /
  3}\left(1+\frac{73}{24} e^{2}+\frac{37}{96}
  e^{4}\right)\left(1-e^{2}\right)^{-7 / 2} \,.
\end{equation}

\subsection{Cubic Galileon} \label{sec:Galileon:theory}

\begin{figure*}
    \includegraphics[width=2\columnwidth]{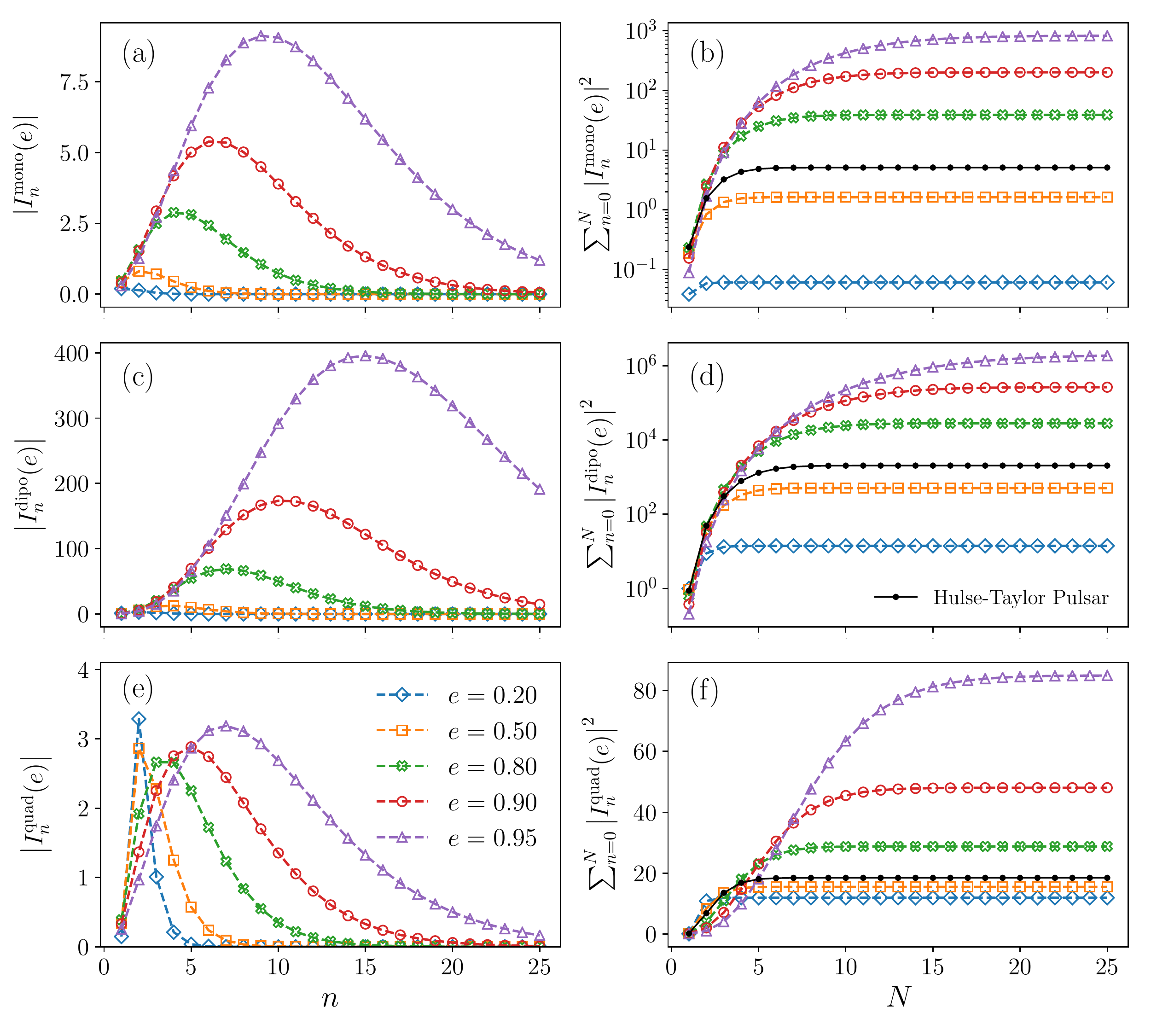}
    \caption{Eccentricity mode functions for the Galileon radiation. The
    left column of the panels gives the absolute values of $I_n^{\rm
    mono}(e)$, $I_n^{\rm dipo}(e)$, and $I_n^{\rm quad} (e)$ from top to
    bottom. The right column of the panels are the cumulative contribution
    of these functions to the Galileon radiation power, in the form of
    $\sum_{n=0}^N \left| I_n^{\rm mono}(e) \right|^2$ for the monopole
    ({\it top}), $\sum_{n=0}^N \left| I_n^{\rm dipo}(e) \right|^2$ for the
    dipole ({\it middle}), and $\sum_{n=0}^N \left| I_n^{\rm quad}
    (e)\right|^2$ for the quadrupole ({\it bottom}). Notice that, for
    clarity panels (b) and (d) have used a logarithmic scale for the
    vertical axes, while all the other axes are using a linear scale.
    \ReplyB{For a direct check, the solid black lines in the right column
    of the panels reproduce the result in Figure 1 of \citet{deRham:2012fw}
    for the Hulse-Taylor pulsar PSR~B1913+16 whose orbital eccentricity is
    $e \simeq 0.617$~\cite{Weisberg:2016jye}.} }
    \label{fig:In}
\end{figure*}

As discussed in the Introduction, (Lorentz invariant) massive gravity
models in the decoupling limit essentially reduce to Galileon models plus
linearized helicity-2 modes. For the cubic Galileon, we focus on the
action~\cite{Luty:2003vm, deRham:2012fw},
\begin{align}\label{eq:cubic:Galileon:action}
    S =\int \mathrm{d}^{4} x \Bigg[& -\frac{1}{4} h^{\mu \nu}(\mathcal{E}
    h)_{\mu \nu} +\frac{h^{\mu \nu} T_{\mu \nu} }{2 M_{\mathrm{Pl}}}
    \nonumber\\
    &~~ -\frac{3}{4}(\partial \pi_s)^{2}\left(1+\frac{1}{3 \Lambda^{3}}
    \square \pi_s \right) + \frac{\pi_s T}{2M_{\rm Pl}} \Bigg] \,,
\end{align}
where $h_{\mu\nu} \equiv g_{\mu\nu}-\eta_{\mu\nu}$ is the perturbation of
the metric, the first two terms in the integrand are the linearized
Einstein-Hilbert term coupled to matter with the Lichnerowicz operator
$(\mathcal{E} h)_{\mu \nu} \equiv -\frac{1}{2} \square h_{\mu \nu}+\cdots$,
$T$ is the trace of the energy-momentum tensor $T^{\mu\nu}$, and $\Lambda$
is the strong coupling scale of the Galileon sector, related to the mass of
graviton $m_g$ via $\Lambda^3 = m_g^2 M_{\rm Pl}$. Therefore, the field
equations for $\pi_s$ and $h_{\mu\nu}$ decouple,
\begin{align}
  \frac{1}{M_{\mathrm{Pl}}} T_{\mu \nu} &= -\frac{1}{2} \square h_{\mu \nu}
  \,, \\
  \frac{1}{2 M_{\mathrm{Pl}}} T &= \partial_{\mu}\left[-\frac{3}{2}
  \partial^{\mu} \pi_s \left(1+\frac{1}{3 \Lambda^{3}} \square
  \pi_s\right)+\frac{1}{4 \Lambda^{3}} \partial^{\mu}(\partial
  \pi_s)^{2}\right]\,.
\end{align}
For a static system whose total mass is $M$, one can define the Vainshtein
radius as,
\begin{equation}
    r_{\star} = \left(\frac{M}{16 m_g^{2} M_{\mathrm{Pl}}^{2}}\right)^{1/3}
    = \frac{1}{\Lambda}\left(\frac{M}{16 M_{\mathrm{Pl}}}\right)^{1/3} \,.
\end{equation}
Within $r_\star$, the fifth force from the scalar degree of freedom is
strongly suppressed.

In a static system, the fifth force is suppressed by a factor $\sim
\left(L/r_\star\right)^{3/2}$, where $L$ is the typical lengthscale of the
system~\cite{Babichev:2013usa}. For example, for an imaginary ``static''
binary system, we can choose it to be the semimajor axis of the orbit, $L
\sim a$. However, binaries are not static. \citet{deRham:2012fw} and
\citet{Chu:2012kz} explicitly worked out the gravitational radiation
behaviors in a time-dependent binary system at lowest orders. They found
that, (i) extra Galileon radiation powers exist at the monopole, dipole,
and quadrupole levels, and (ii) for the dominant radiation the suppression
factor is weakened from $\left(L/r_\star\right)^{3/2}$ to $\left(n_b
r_\star\right)^{-3/2}$.

Similar to GR, the Newtonian-order contribution of both the monopole and
dipole Galileon radiation vanishes, due to the conservation of energy and
linear momentum of the system respectively. Post-Newtonian order
contributions do exist for them. For the quadrupole radiation, the
Newtonian-order contribution is nonzero. We collect the radiation powers of
these multiple moments from Ref.~\cite{deRham:2012fw} \ReplyB{(see also the
numerical calculation in Ref.~\cite{Dar:2018dra})} in the following for
later use in this work.
\begin{itemize}
  \item {\it Monopole radiation}. In the cubic Galileon
  model~(\ref{eq:cubic:Galileon:action}), the monopole radiation power at
  the leading order for a binary system is,
    \begin{equation}
        \label{eq:power:mono}
        {\cal P}_{\rm mono} = \frac{25}{48} \pi\beta^{2} n_b^{2}
        \frac{\left(n_b a \right)^{4}}{\left(n_b r_{\star}\right)^{3 / 2}}
        \frac{ M_{\rm mono}^2}{M_{\mathrm{Pl}}^{2}}
        \sum_{n=0}^\infty \left|I_{n}^{\rm mono}(e)\right|^{2} \,,
    \end{equation}
  where the constant $\beta$ and the ``monopole mass'' $M_{\rm mono}$ (also
  known as the reduced mass) are defined respectively as,
  \begin{equation}
    \beta \equiv \frac{3^{3/8}}{\Gamma(3/4)} \left( \frac{\pi}{32}
    \right)^{1/4} \simeq 0.6897 \,,
  \end{equation}
  \begin{equation}
    M_{\rm mono} \equiv \eta M = \frac{m_1 m_2}{M} \,,
  \end{equation}
  with $\Gamma(\cdot)$ the gamma function, and the eccentricity mode
  function $I^{\rm mono}_n(e)$ is given in Eq.~(\ref{eq:In:mono}) and
  Fig.~\ref{fig:In}.
  \item {\it Dipole radiation}. The dipole radiation power at the leading
  order for a binary system is,
    \begin{equation}
        \label{eq:power:dipo}
        {\cal P}_{\text { dipole }}=\frac{c_{1}^{2}}{8} n_b^{2}
        \frac{\left(n_b a\right)^{6}}{\left(n_b r_{\star}\right)^{3 / 2}}
        \frac{M_{\rm dipo}^{2}}{M_{\mathrm{Pl}}^{2}}
        \sum_{n=0}^{\infty}\left|I_{n}^{\rm dipo}(e)\right|^{2} \,,
    \end{equation}
  where the constant $c_1$ and the ``dipole mass'' are defined respectively
  as ,
    \begin{equation}
        c_1 \equiv \frac{3^{7/8} \left(\pi/2\right)^{1/4}}{8\,\Gamma(7/4)}
        \left[ 1 + \frac{3}{16} \frac{\Gamma(7/4)}{\Gamma(11 / 4)} \right]
        \simeq 0.4408 \,,
    \end{equation}
    \begin{equation}
        M_{\rm dipo} \equiv {\cal X} M_{\rm mono} = M_{\rm mono} \left
        (\frac{m_1-m_2}{M}\right) \,,
    \end{equation}
  where ${\cal X} \equiv \left(m_1 - m_2\right)/M$, and the eccentricity
  mode function $I_{n}^{\rm dipo}(e)$ is given in Eq.~(\ref{eq:In:dipo})
  and Fig.~\ref{fig:In}.
  \item {\it Quadrupole radiation}. Besides the quadrupole radiation power
  in Eq.~(\ref{eq:power:GR}), the extra power at the leading order reads,
    \begin{equation}
        \label{eq:power:quad}
        {\cal P}_{\rm quad}=\frac{5 \lambda^{2}}{32} n_b^{2}\frac{\left(n_b
        a\right)^{3}}{\left(n_b r_{\star}\right)^{3 / 2}} \frac{M_{\rm
        quad}^{2}}{M_{\mathrm{Pl}}^{2}} \sum_{n=0}^{\infty}\left|I_{n}^{\rm
        quad}(e)\right|^{2} \,,
    \end{equation}
  where the constant $\lambda$ and the ``quadrupole mass'' $M_{\rm quad}$
  are defined respectively as,
    \begin{equation}
        \lambda \equiv \frac{3^{9 / 8} \pi^{1 / 4}}{2^{17 / 4} \Gamma(9 /
        4)} \simeq 0.2125 \,,
    \end{equation}
    \begin{equation}
        M_{\rm quad} \equiv {\cal Y} M_{\rm mono} = M_{\rm mono}
        \left(\frac{\sqrt{m_{1}}+\sqrt{m_{2}}}{\sqrt{M}}\right) \,,
    \end{equation}
  where ${\cal Y} \equiv \left(\sqrt{m_1} + \sqrt{m_2}\right) / \sqrt{M} $,
  and the eccentricity mode function $I_{n}^{\rm quad}(e)$ is given in
  Eq.~(\ref{eq:In:quad}) and Fig.~\ref{fig:In}.
\end{itemize}

The eccentricity mode functions, $I_{n}^{\rm mono}(e)$, $I_{n}^{\rm
dipo}(e)$, and $I_{n}^{\rm quad}(e)$, can be defined in a uniform way via
the master function,
\begin{equation}
    I_n^{\left( p,q \right)} (e) \equiv \frac{n^{ p + \frac{1}{4}}}{2\pi}
    \left(1-e^2\right)^p \int_{0}^{2\pi} \frac{ e^{-i q x}}{ \left(1 + e
    \cos x \right)^p } {\rm d} x \,.
\end{equation}
The above mentioned eccentricity functions for monopole, dipole, and
quadrupole radiations are,
\begin{eqnarray}
    \label{eq:In:mono} I_n^{\rm mono} (e) &=& I_n^{ \left( p = 2,
    q = n \right)} (e) \,, \\
    \label{eq:In:dipo} I_n^{\rm dipo} (e) &=& I_n^{ \left( p = 3,
    q = n-1 \right) } (e) \,, \\
    \label{eq:In:quad} I_n^{\rm quad} (e) &=& I_n^{ \left( p = 3/2,
    q = n-2 \right)} (e) \,.
\end{eqnarray}
The behaviors of these functions are illustrated in Fig.~\ref{fig:In} for
different values of the eccentricity.

Using the same reasoning of energy balance in Sec.~\ref{sec:GR}, we can get
the extra contributions to $\dot P_b$ from the extra Galileon radiation
powers in Eqs.~\eqref{eq:power:mono}, \eqref{eq:power:dipo}, and
\eqref{eq:power:quad},
\begin{align}
  \dot P_b^{\rm mono} &= -25\sqrt{2}\beta^2 \pi^{5/2} \frac{\eta (GM)^{7/6}
  n_b^{1/6}}{\hbar c^{3/2}} \,m_g \sum_{n=0}^{\infty}\left|I_{n}^{\rm
  mono}(e)\right|^{2} \,, \label{eq:Pbdot:mono} \\
  \dot P_b^{\rm dipo} &= - 6\sqrt{2} c_1^2 \pi^{3/2} \frac{ \eta {\cal X}^2
  (GM)^{11/6} n_b^{5/6} }{ \hbar c^{7/2}} \,m_g
  \sum_{n=0}^{\infty}\left|I_{n}^{\rm dipo}(e)\right|^{2} \,,
  \label{eq:Pbdot:dipo} \\
  \dot P_b^{\rm quad} &= - \frac{15}{\sqrt{2}} \lambda^2 \pi^{3/2}
  \frac{{\eta \cal Y}^2 (GM)^{5/6}}{n_b^{1/6} \hbar c^{1/2} } \,m_g
  \sum_{n=0}^{\infty}\left|I_{n}^{\rm quad}(e)\right|^{2} \,.
  \label{eq:Pbdot:quad}
\end{align}
\ReplyA{In obtaining these results, at the leading order we have used the
Newtonian binding energy for the orbit in Eq.~(\ref{eq:orbit:energy}), and
we have made use of the orbital-averaged radiating powers as discussed
above. Such a simplification is sufficient for the analysis in this paper.}
Notice that the extra contributions to $\dot P_b$ are all proportional to
the graviton mass $m_g$, while in the linearized Fierz-Pauli
theory~\cite{Finn:2001qi, Seymour:2018bce, Miao:2019nhf}, the extra $\dot
P_b \propto m_g^2$.

\section{Constraints from binary pulsars} \label{sec:psr}

Now we would like to better understand the physical effect of the Galileon
radiation in Eqs.~(\ref{eq:Pbdot:mono}--\ref{eq:Pbdot:quad}) to different
kinds of binary pulsars. We present some general consideration in
Sec.~\ref{sec:psr:general}, concerning the dependence on the orbital
eccentricity $e$, the orbital period $P_b$, and the component masses
$\left(m_1, m_2\right)$. We cast constraints on the graviton mass in the
cubic Galileon theory from precision timing of fourteen carefully-chosen
binary pulsars in Sec.~\ref{sec:psr:constraint}.

\subsection{General consideration} \label{sec:psr:general}

From Fig.~\ref{fig:In} we see that, (i) given an eccentricity, the absolute
values of the eccentricity mode functions, $I_n(e)$, increase at first,
then have a peak at a particular $n = n^{\rm peak}$ before they decrease;
(ii) with the eccentricity increasing, $n^{\rm peak}$ happens at a larger
$n$, therefore when we deal with highly eccentric binary pulsars, more
modes are to be included in order to guarantee the convergence of the sum;
and (iii) the cumulative contributions of these mode functions to the
Galileon radiation powers in Eqs.~\eqref{eq:power:mono},
\eqref{eq:power:dipo}, and \eqref{eq:power:quad} saturate after a
particular $n$, and for $e \lesssim 0.95$, summing up to $N \sim 15$ is
generally sufficient. In our following calculation, we choose $N=30$ for a
better accuracy. However, one should remember that, for extremely eccentric
binaries with $1-e \ll 10^{-2}$, a larger cutoff $N$ is needed.

\begin{figure}[t]
  \includegraphics[width=8.8cm]{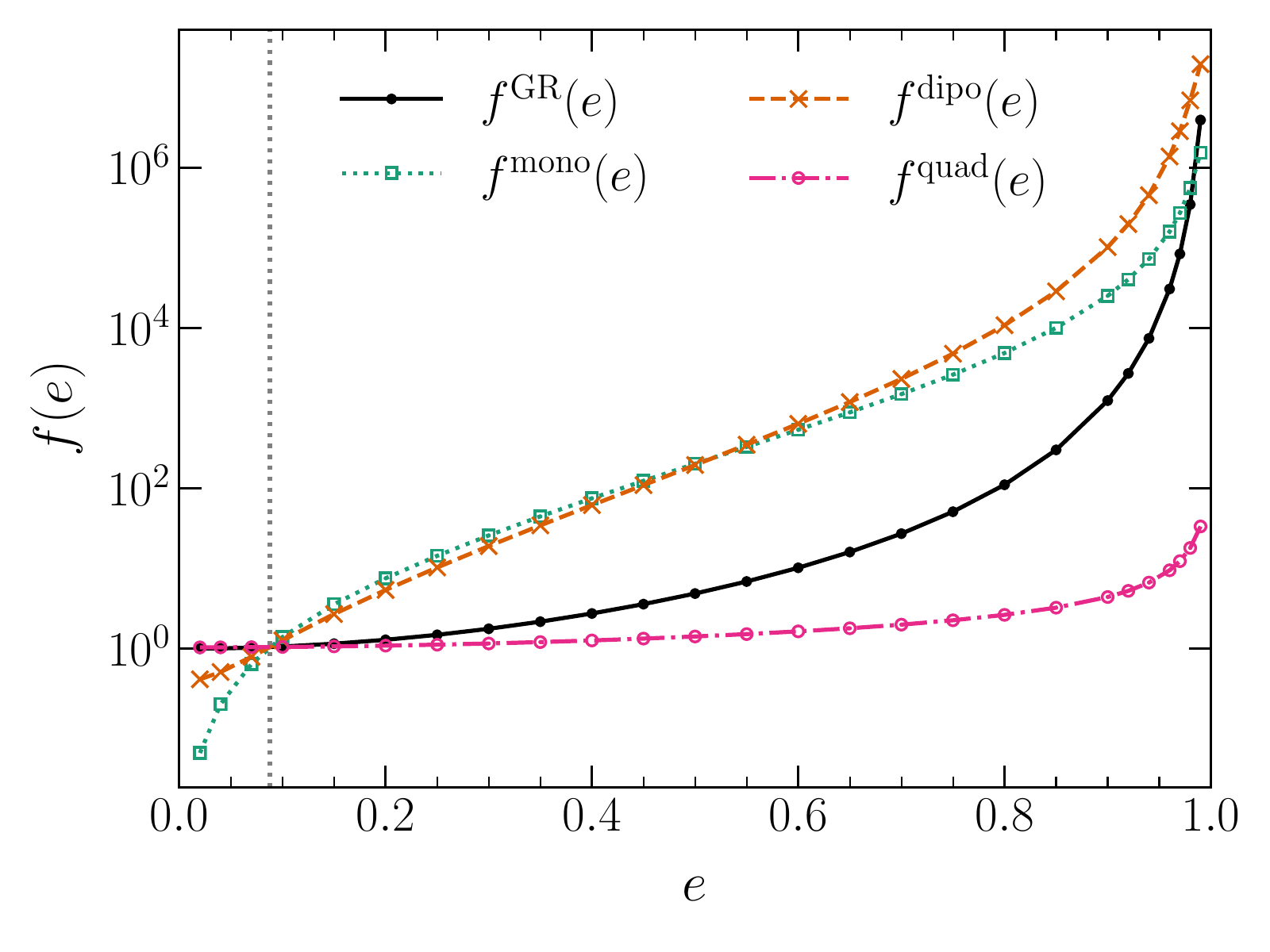}
  \caption{\label{fig:ecc} The $f(e)$ factor in $\dot P_b$, defined in
  Eqs.~(\ref{eq:fe:GR}--\ref{eq:fe:quad}), with a normalization such that
  $f^{\rm GR} = f^{\rm mono} = f^{\rm dipo} = f^{\rm quad}$ for the double
  pulsar~\cite{Kramer:2006nb} whose $e\simeq 0.088$ (dotted vertical
  line).}
\end{figure}

Based on the contribution to the Galileon radiation power, we define,
\begin{align}
  f^{\rm GR}(e) &\equiv \left(1+\frac{73}{24} e^{2}+\frac{37}{96}
  e^{4}\right)\left(1-e^{2}\right)^{-7 / 2} \,,  \label{eq:fe:GR} 
\end{align}
and
\begin{align}
  f^{\rm mono}(e) &\propto \sum_{n=0}^{\infty}\left|I_{n}^{\rm mono}(e)\right|^{2}
  \,, \label{eq:fe:mono} \\
  f^{\rm dipo}(e) &\propto \sum_{n=0}^{\infty}\left|I_{n}^{\rm
  dipo}(e)\right|^{2} \,, \label{eq:fe:dipole} \\
  f^{\rm quad}(e) &\propto \sum_{n=0}^{\infty}\left|I_{n}^{\rm
  quad}(e)\right|^{2} \,. \label{eq:fe:quad}
\end{align}
These are the dependence shown in the GR radiation (\ref{eq:Pbdot:GR}) and
Galileon radiation (\ref{eq:Pbdot:mono}--\ref{eq:Pbdot:quad}). The
proportional factors in Eqs.~(\ref{eq:fe:mono}--\ref{eq:fe:quad}) can be
arbitrary normalization values for the convenience of specific demonstration.

The $f(e)$ functions in Eqs.~(\ref{eq:fe:GR}--\ref{eq:fe:quad}) are plotted
in Fig.~\ref{fig:ecc} with a choice of normalization. We can see that the
more eccentric binaries are emitting more gravitational waves and
contributing more significantly to $\dot P_b$. This is true for the three
types of Galileon radiation, as well as the quadrupole radiation in GR. If
the binary is extremely eccentric with $1-e \ll 1$, a huge amplificative
factor occurs, especially for the dipole radiation (\ref{eq:Pbdot:dipo}) in
the cubic Galileon. As we will see below, the quadrupole Galileon radiation
is in general the main contributor to $\dot P_b$ for binaries with
comparable component masses~\cite{deRham:2012fw}. The increase of $f^{\rm
quad}(e)$ at large eccentricities is slower than the other ones. Notice
that the curves in the figure only represent {\it one} numerical factor in
the radiation power, defined in Eqs.~(\ref{eq:fe:GR}--\ref{eq:fe:quad}),
while the other dependences (for example, the dependence on the orbital
period and the component masses) are omitted here, which will be
investigated in the following. \ReplyB{Worth to note that, generally the
increased radiation due to eccentricity will tend to circularize the
orbits (see e.g. in GR~\cite{Peters:1963ux}), making highly eccentric
binaries less likely to exist to date.}

\begin{figure}[t]
    \includegraphics[width=8.8cm]{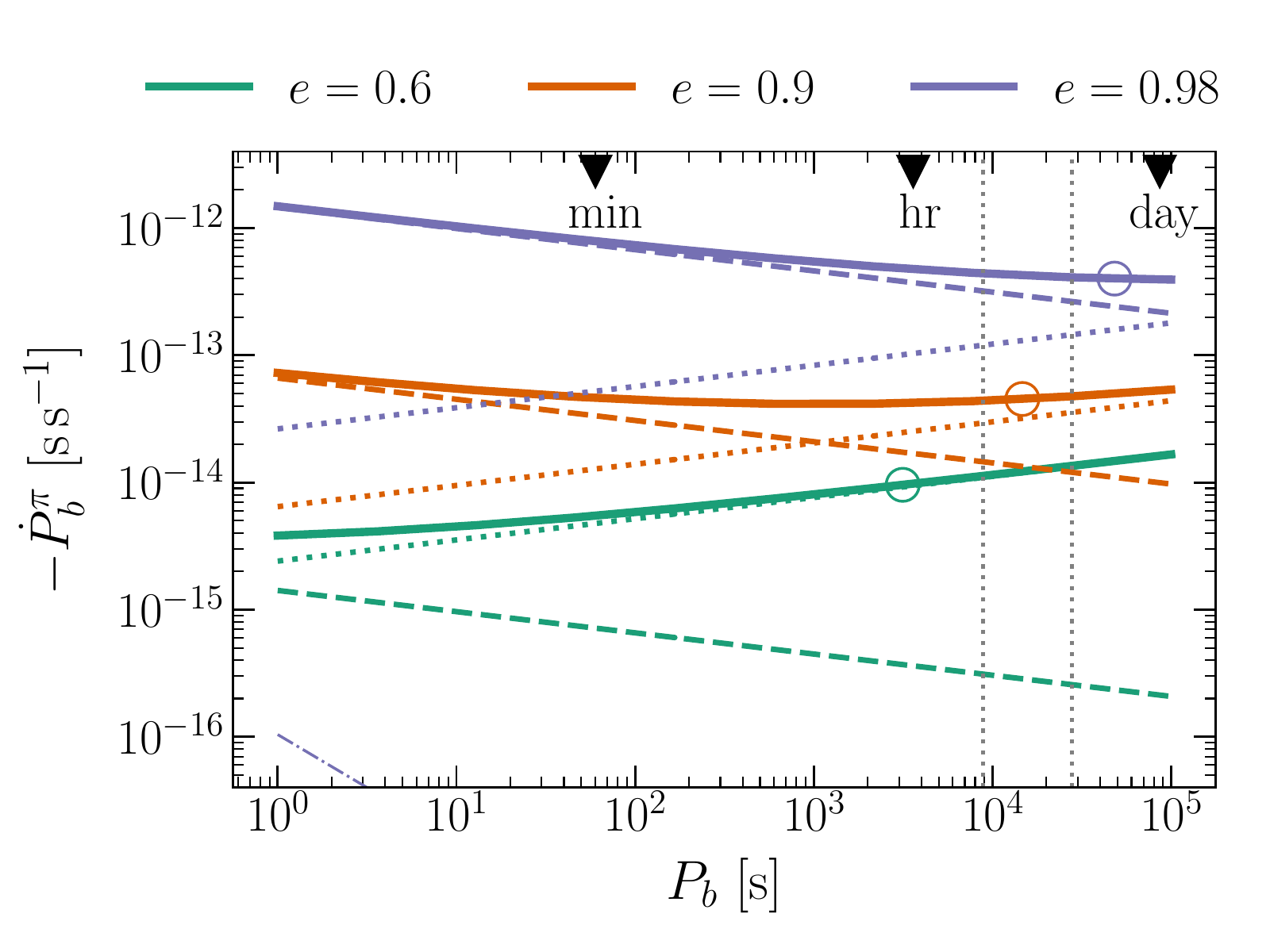}
    \caption{\label{fig:Pb} Galileon radiation induced orbital decay as a
    function of the orbital period $P_b$, for three different eccentricities.
    We have assumed the graviton mass $m_g = 10^{-27}\, {\rm eV}/c^2$, and
    component masses as per the Hulse-Taylor pulsar, namely $m_1 =
    1.438\,M_\odot$ and $m_2 = 1.390\,M_\odot$. The dotted vertical lines
    indicate the orbital periods of the double pulsar ({\it left}) and the
    Hulse-Taylor pulsar ({\it right}). The dashed, dot-dashed, dotted lines
    indicate contributions from the monopole, dipole, and quadrupole
    radiations respectively, while the solid lines are their sum. The dipole
    radiation is only visible for $e=0.98$ at the bottom left corner of the
    figure. Open circles on the solid curves represent systems that
    have lifetime of 1\,Myr before merger in GR.}
\end{figure}

It is interesting to observe that, compared with $\dot P_b^{\rm GR}/P_b
\propto n_b^{8/3}$ in GR [see Eq.~\eqref{eq:Pbdot:GR}], Galileon radiations
have $\dot P_b^{\rm mono} / P_b \propto n_b^{7/6}$ [see
Eq.~\eqref{eq:Pbdot:mono}], $\dot P_b^{\rm dipo}/P_b \propto n_b^{11/6}$
[see Eq.~\eqref{eq:Pbdot:dipo}], and $\dot P_b^{\rm quad}/P_b \propto
n_b^{5/6}$ [see Eq.~\eqref{eq:Pbdot:quad}]. It is a noteworthy feature for
the gravitational radiation with the screening mechanism. With such a
dependence, while more relativistic binaries are more prominent in emitting
gravitational waves in GR, this may not always be true in the cubic
Galileon. It resembles the dependence on $n_b$ in Lorentz-violating massive
gravity~\cite{Finn:2001qi, Miao:2019nhf} and gravity theories with a
time-varying gravitational ``constant'' $G(t)$~\cite{Wex:2014nva,
Seymour:2018bce}.

In Fig.~\ref{fig:Pb}, we plot three types of Galileon radiation, as
function of the orbital period, for a binary pulsar system with component
masses similar to the Hulse-Taylor pulsar, and a graviton mass $m_g =
10^{-27}\,{\rm eV}/c^2$. First, we observe that, the dipole radiation is
orders of magnitude smaller than the other two types of Galileon radiation,
thus it can be totally ignored. Though, as we recall from
Fig.~\ref{fig:ecc}, the dependence of the $f(e)$ factor for the dipole
radiation on the eccentricity $e$ is one of the steepest, the overall
effect from dipole radiation is negligible even for $e=0.98$. Then, we
observe in Fig.~\ref{fig:Pb} that, while for binaries with $e \lesssim
0.6$, the quadrupole radiation dominates~\cite{deRham:2012fw}, for highly
eccentric binaries with relativistic orbits (namely, smaller $P_b$), the
monopole radiation might dominate instead. This happens when $P_b \lesssim
10^3\,{\rm s}$ for $e=0.9$, and $P_b \lesssim 10^5\,{\rm s}$ for $e=0.98$.
It is caused by the facts that $\dot P_b^{\rm mono}$ increases for larger
$n_b$ (smaller $P_b$), while $\dot P_b^{\rm quad}$ increases for smaller
$n_b$ (larger $P_b$). Therefore, in particular for highly eccentric
binaries to be discovered in the future, we should keep in caution of only
considering the quadrupole Galileon radiation. This also applies to highly
eccentric Galactic binaries (if they exist) in the mHz gravitational-wave
band for the Laser Interferometer Space Antenna
(LISA)~\cite{Audley:2017drz, Barausse:2020rsu}. A similar discussion will
also be mentioned for pulsar-BH systems in Sec.~\ref{sec:bhs}. In our
following calculation, we keep all three types of Galileon radiation
summed, no matter of their relative strength.

\begin{figure}[t]
  \includegraphics[width=8.8cm]{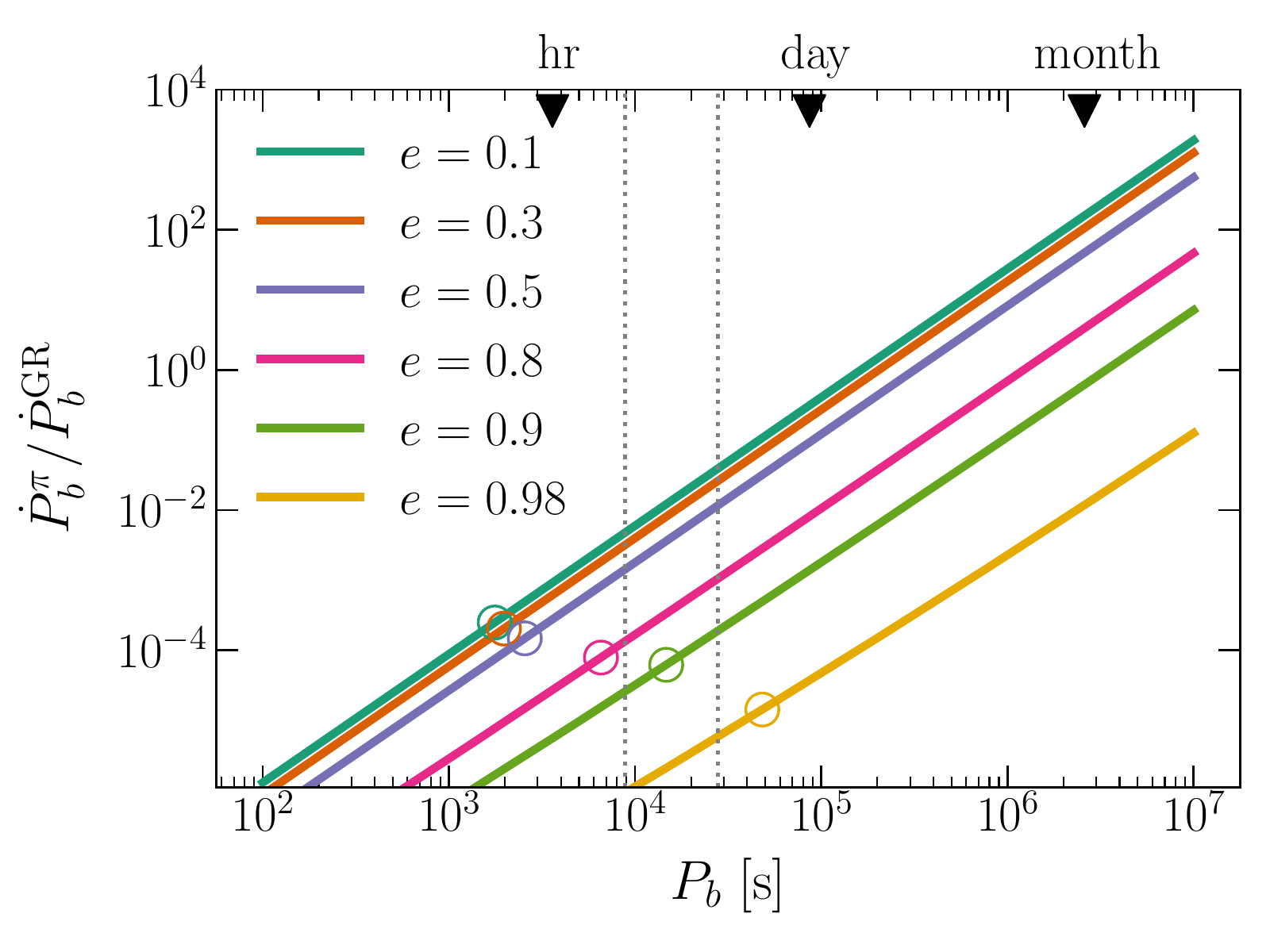}
  \caption{\label{fig:Pb:GR} The contribution from the Galileon radiation
  (summation of the monopole, dipole, and quadrupole radiations), relative
  to the quadrupole radiation in GR, for several different eccentricities.
  We have assumed $m_1$, $m_2$, and $m_g$, as per Fig.~\ref{fig:Pb}. The
  dotted vertical lines indicate the orbital periods for the double pulsar
  ({\it left}) and the Hulse-Taylor pulsar ({\it right}). Open circles
  represent systems that have lifetime of 1\,Myr before merger in GR.}
\end{figure}

In Fig.~\ref{fig:Pb:GR}, we plot the relative contribution of the Galileon
radiation to the quadrupole radiation in GR. First of all, we see that the
relative contribution increases when the orbital period is larger. This is
in accordance with the screening mechanism which works less well when the
size increases~\cite{Babichev:2013usa}, while the GR effects are more
prominent when the orbit is more compact. Specifically it is caused by
that, (i) in this regime, the quadrupole Galileon radiation dominates,
which increases as $\propto P_b^{1/6}$ when $P_b$ becomes larger; (ii) the
quadrupole radiation in GR decreases as $\propto P_b^{-5/3}$. With $m_g =
10^{-27} \,{\rm eV}/c^2$ the Galileon radiation may even be larger than the
GR one when $P_b \gtrsim {\rm day}$ for nearly circular orbits. Moreover,
we find that, while the absolute Galileon quadrupole radiation increases
with larger eccentricity (see Fig.~\ref{fig:ecc}), the relative
contribution to GR decreases. It can be understood from the steepness of
the curves in Fig.~\ref{fig:ecc} for the quadrupole radiations in GR (black
dots) and in the cubic Galileon (magenta circles). The latter one is less
steep, thus decreasing the relative ratio of $\dot P_b$ with larger
eccentricity.

\begin{figure*}
  \includegraphics[width=1.8\columnwidth]{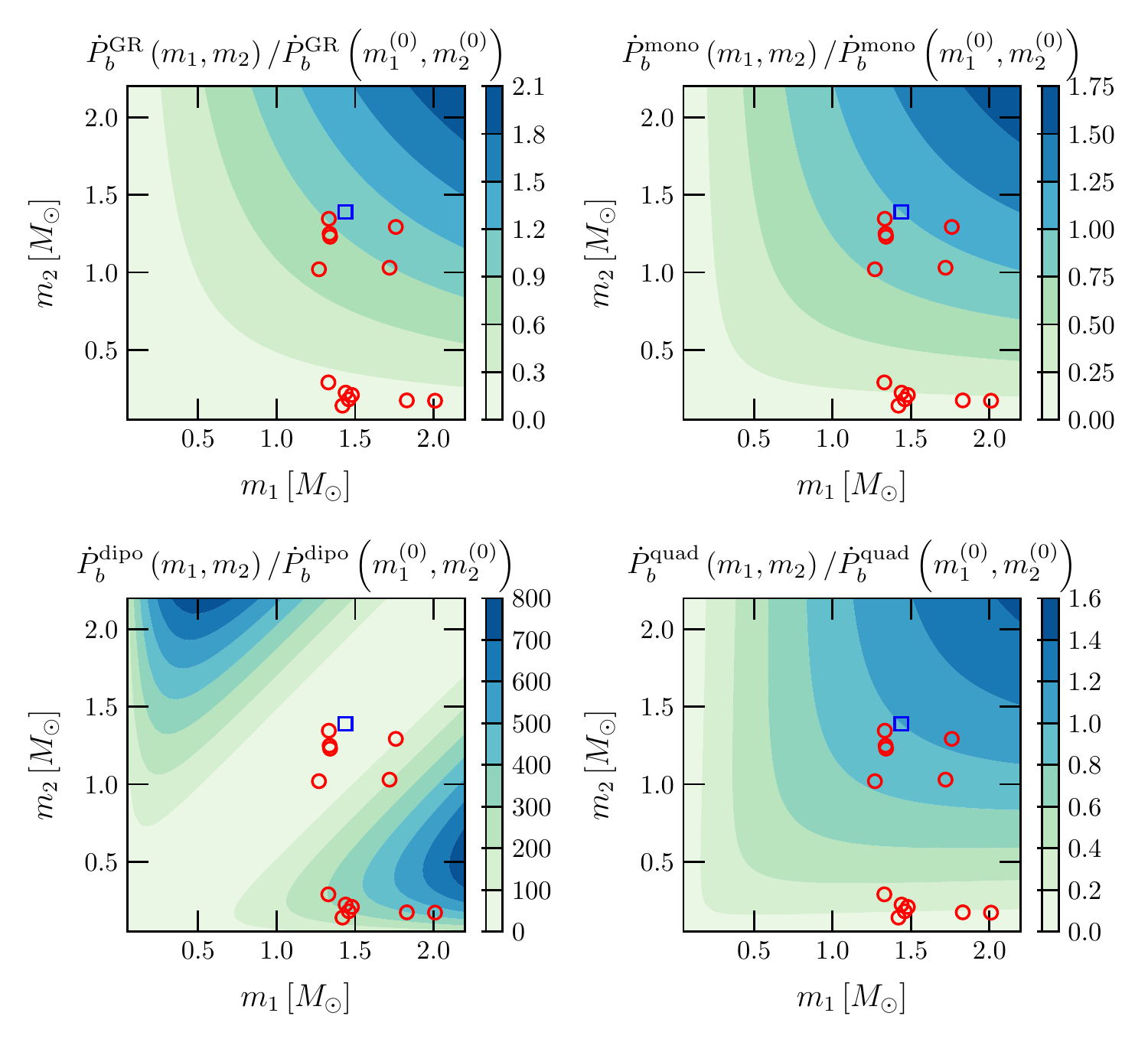}
  \caption{The relative strength to $\dot P_b$ versus component masses
  $\left(m_1, m_2\right)$, with given orbital period $P_b$ and orbital
  eccentricity $e$. The panels are for the GR radiation ({\it top left}),
  and monopole ({\it top right}), dipole ({\it lower left}), quadrupole
  ({\it lower right}) Galileon radiations. The results are calculated with
  respect to a fiducial mass pair, $m_1 = 1.438\,M_\odot$ and $m_2 =
  1.390\,M_\odot$, denoted by the blue square. Red circles are the other
  thirteen pulsars that are used in this paper (see Table~\ref{tab:psr}).}
  \label{fig:contour:mass}
\end{figure*}

\ReplyB{Lastly, to understand the effect of component masses, in
Fig.~\ref{fig:contour:mass} we plot the relative strength to $\dot P_b$ for
different mass pairs, normalized to $\left(m_1,m_2\right) = (1.438,
1.390)\,M_\odot$, which is the fiducial mass pair based on the
Hulse-Taylor pulsar~\cite{Weisberg:2016jye}.} This figure is useful to
analyze binaries with given orbital period and eccentricity, but different
component masses. It is well known that, for the quadrupole radiation in
GR, the so-called chirp mass, ${\cal M} \equiv \left(m_1m_2\right)^{3/5} /
M^{1/5}$, plays the key role, whose contours in the $m_1$-$m_2$ parameter
space are shown in the upper left panel in Fig.~\ref{fig:contour:mass}. The
monopole and quadrupole Galileon radiations depend on the component masses
in a quantitatively different, but qualitatively similar way, as shown in
the right column of panels. But for the dipole radiation, shown in the
bottom left panel, the dependence is totally different. Asymmetric binaries
are preferred to manifest the dipole radiation. Notice that from the
colorbars, the enhancement from asymmetric component masses for the dipole
radiation can be as large as $\sim 10^3$, while for the other three types
of radiation, the change with component masses is within a relatively
limited range. This point will be clearly observed for pulsar-BH systems in
Sec.~\ref{sec:bhs}.

\subsection{Constraints} \label{sec:psr:constraint}

\def\arraystretch{1.5}
\begin{table*}[t]
  \caption{Relevant parameters for a collection of binary pulsars to test
  the Galileon radiation. Component masses were derived using
  post-Keplerian parameters other than $\dot P_b$, based on the validity of
  GR; $\sigma_{\dot P_b^{\rm obs}}$ is the measurement uncertainty of the
  timing parameter $\dot P_b$, while $\sigma_{\dot P_b^{\rm int}}$ is the
  uncertainty of $\dot P_b$ after accounting for the kinematic Shklovskii
  effect and the Galactic acceleration
  contribution~\cite{Damour:1990wz,Lorimer:2005misc}. Parenthesized numbers
  represent the 1-$\sigma$ uncertainty in the last digit(s) quoted.
  \ReplyB{For more details, interested readers are referred to the cited
  references alongside the pulsar name for each pulsar.}
  \label{tab:psr}}
\centering
\begin{tabular}{p{3cm}p{3cm}p{2.5cm}p{1.7cm}p{1.7cm}p{2.2cm}p{2.2cm}}
\hline\hline
Pulsar & $P_b \,[{\rm d}]$ & $e$ & $m_1\,[M_\odot]$ & $m_2\,[M_\odot]$ & $ \sigma_{\dot P_b^{\rm obs}} \, [{\rm s\,s}^{-1}]$ & $\sigma_{\dot P_b^{\rm int}} \, [{\rm s\,s}^{-1}]$ \\
\hline
J0348+0432~\cite{Antoniadis:2013pzd} & 0.102424062722(7) & $(2.6\pm0.9)\times10^{-6}$ & 2.01(4) & 0.172(3) & $4.5\times10^{-14}$ & $4.5\times10^{-14}$ \\
J0437$-$4715~\cite{Reardon:2015kba,Perera:2019sca} & 5.7410458(3) & $1.9182(1)\times10^{-5}$ & 1.44(7) & 0.224(7) & $3\times10^{-15}$ & $2.8\times10^{-13}$ \\
J0613$-$0200~\cite{Desvignes:2016yex,Perera:2019sca} & 1.198512575217(10) & $4.50(9)\times10^{-6}$ & 1.42(46) & 0.14(3) & $7\times10^{-15}$ & $2.3\times10^{-14}$ \\
J0737$-$3039A~\cite{Kramer:2006nb, Kramer:2016kwa} & 0.10225156248(5) & 0.0877775(9) & 1.3381(7) & 1.2489(7) & $1\times10^{-15}$ & $1\times10^{-15}$ \\
J1012+5307~\cite{Lazaridis:2009kq,Perera:2019sca} &  0.604672723085(3) & $(1.1\pm0.1)\times10^{-6}$ & 1.83(11) & 0.174(7) & $4\times10^{-15}$ & $8\times10^{-15}$ \\
J1022+1001~\cite{Desvignes:2016yex, Perera:2019sca} & 7.805136(1) & $9.704(5) \times10^{-5}$ & 1.72(65) & 1.03(36) & $7\times10^{-14}$ & $2\times10^{-13}$  \\
J1141$-$6545~\cite{Bhat:2008ck, Krishnan:2020txo} & 0.19765096149(3) & 0.171876(1) & 1.27(1) & 1.02(1) & $2.5\times10^{-14}$ & $2.5\times10^{-14}$ \\
B1534+12~\cite{Fonseca:2014qla} & 0.420737298879(2) & 0.27367752(7) & 1.3330(2) & 1.3455(2) & $3\times10^{-16}$ & $1.1\times10^{-14}$ \\
J1713+0747~\cite{Zhu:2018etc, Perera:2019sca} & 67.8251299228(5) & $7.49403(7) \times 10^{-5}$  & 1.33(10) & 0.290(11) & $1\times10^{-13}$ & $1\times10^{-13}$ \\
J1738+0333~\cite{Freire:2012mg} & 0.3547907398724(13) & $(3.4 \pm 1.1) \times 10^{-7}$ & $1.46(6)$ & $0.181(8)$ & $3.1\times10^{-15}$ & $3.2\times10^{-15}$  \\
J1756$-$2251~\cite{Ferdman:2014rna} & 0.31963390143(3) & 0.1805694(2) & 1.341(7) & 1.230(7) &  $5\times10^{-15}$ & $8\times10^{-15}$ \\
J1909$-$3744~\cite{Desvignes:2016yex, Perera:2019sca} & 1.533449475278(1) & $1.04(6)\times10^{-7}$ & 1.48(3) & 0.209(1) & $3\times10^{-15}$ & $1.4\times10^{-14}$ \\
B1913+16~\cite{Weisberg:2016jye} & 0.322997448918(3) & 0.6171340(4) & 1.438(1) & 1.390(1) & $1\times10^{-15}$ & $4\times10^{-15}$ \\
J2222$-$0137~\cite{Cognard:2017xyr} & 2.44576456(13) & $0.380940(3)\times10^{-4}$ & 1.76(6) & 1.293(25) & $9\times10^{-14}$ & $9\times10^{-14}$ \\
\hline
\end{tabular}
\end{table*}

After having looked at the general features for the Galileon radiation in
the last subsection, now we turn to realistic binary pulsar systems. We
carefully choose fourteen binary pulsars that are known to be precisely
timed. They are chosen from the pool of millisecond pulsars in the second
Data Release of the International Pulsar Timing Array
program~\cite{Perera:2019sca}, as well as several other well-known systems
with measurement of $\dot P_b$~\cite{Antoniadis:2013pzd, Reardon:2015kba,
Desvignes:2016yex, Kramer:2006nb, Kramer:2016kwa, Lazaridis:2009kq,
Bhat:2008ck, Krishnan:2020txo, Fonseca:2014qla, Zhu:2018etc, Freire:2012mg,
Ferdman:2014rna, Weisberg:2016jye, Cognard:2017xyr}. The relevant
parameters for the test of the Galileon radiation are listed in
Table~\ref{tab:psr}. Because binary pulsars have a large variety in terms
of system parameters and observational characteristics, for detailed
description of these systems, readers are referred to the original timing
papers which are given for each pulsar in the table. \ReplyB{As one can see
from the last subsection, the Galileon radiation depends on a set of physical
parameters of the binary pulsar system. These pulsars in
Table~\ref{tab:psr} represent, to our best knowledge and resource, a set of
the most suitable binary pulsar systems to date to perform the test in this
paper.}

For the chosen binary pulsars, post-Keplerian parameters, other than $\dot
P_b$, were used to calculate the component masses $m_1$ and $m_2$, assuming
the validity of GR expressions~\cite{Lorimer:2005misc,Wex:2014nva}. The
component masses are listed in the fourth and fifth columns in
Table~\ref{tab:psr}. In the strictest sense, the component masses should be
calculated consistently using cubic Galileon theory instead of GR.
Nevertheless, as the Vainshtein suppression of the fifth force is more
significant than the Galileon radiation~\cite{deRham:2012fw,
deRham:2012fg}, it is safe to use the GR formulae to extract these
parameters.

As we can see in the penultimate column in the table, most of the chosen
pulsars have uncertainties for the value of measured $\dot P_b$ at the
level of $\sigma_{\dot P_b^{\rm obs}} \sim 10^{-15} \, {\rm s\,s}^{-1}$.
PSR~B1534+12 has an even better measurement with $\sigma_{\dot P_b^{\rm
obs}} \simeq 3 \times 10^{-16} \, {\rm s\,s}^{-1}$~\cite{Fonseca:2014qla}.
The excellent timing precision is attributed to the long-term observation
of these pulsars and the continuous improvements of the instruments at
large radio telescopes~\cite{Lorimer:2005misc, Kramer:2004hd,
Shao:2014wja}. However, these values cannot be directly used due to the
astrophysical contribution and imperfect knowledge about their distances
as well as the Milky Way's gravitational potential~\cite{Damour:1990wz,
Lorimer:2005misc}. The most significant contribution comes from the
kinematic Shklovskii effect and the Galactic acceleration contribution.
These contributions need to be subtracted using the measurement of the
proper motion and the modeling of the Galactic
potential~\cite{Damour:1990wz}. The subtraction introduces extra
uncertainties in the intrinsic $\dot P_b$ parameter. The uncertainty after
the subtraction, denoted as $\sigma_{\dot P_b^{\rm int}}$, is listed in the
last column of Table~\ref{tab:psr} for each pulsar.

\begin{figure*}
  \includegraphics[width=1.8\columnwidth]{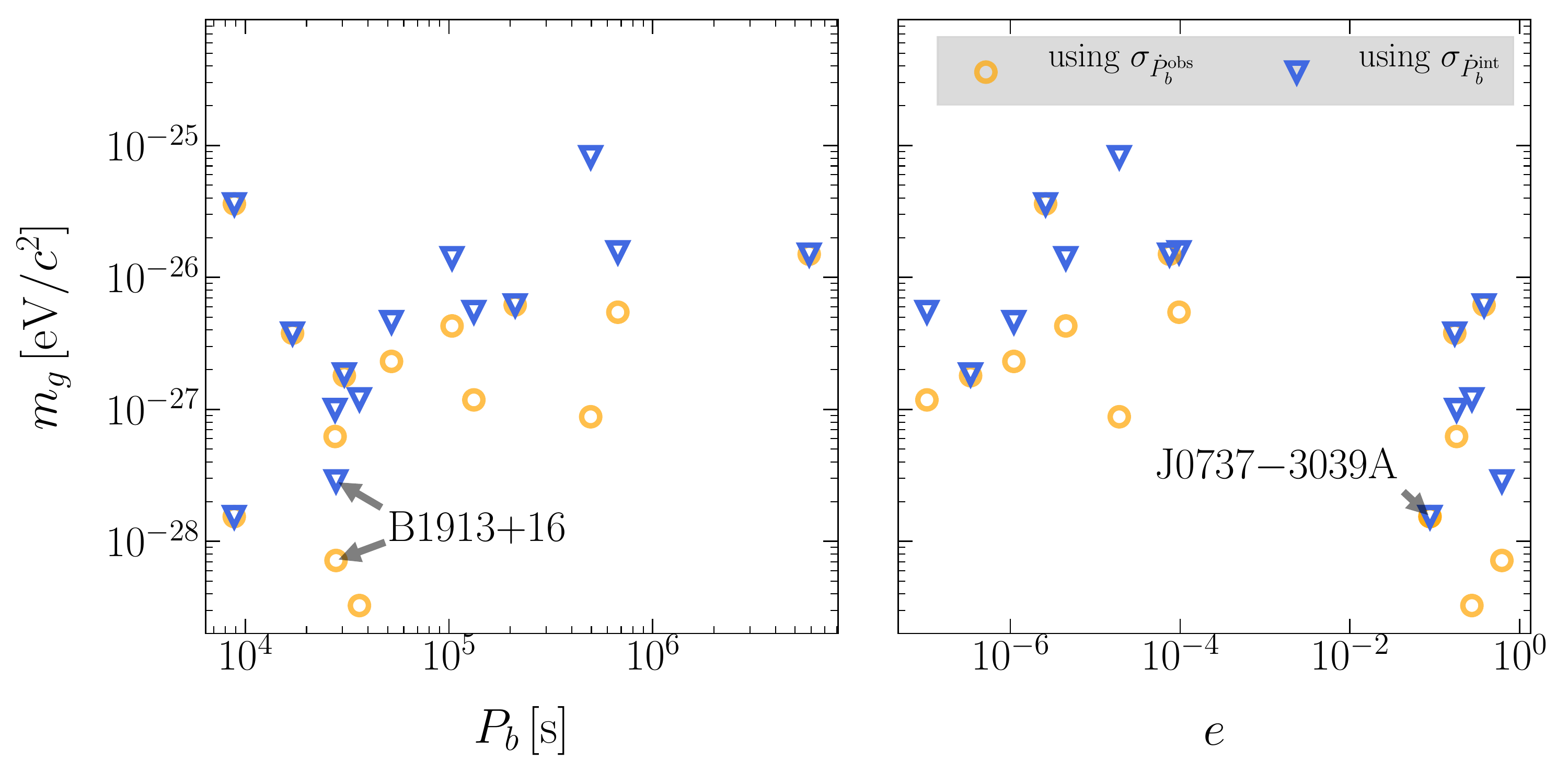}
  \caption{Constraints on the graviton mass in cubic Galileon theory from
  individual binary pulsars versus their orbital periods ({\it left}) and
  eccentricities ({\it right}) \ReplyB{at 68\% C.L.}. Orange circles and
  blue triangles are bounds obtained using $\sigma_{\dot P_b^{\rm obs}}$
  and $\sigma_{\dot P_b^{\rm int}}$, respectively. The Hulse-Taylor pulsar
  PSR~B1913+16~\cite{Weisberg:2016jye} and the double pulsar
  PSR~J0737$-$3039A~\cite{Kramer:2016kwa} are annotated.}
  \label{fig:constraint}
\end{figure*}

\begin{figure}
  \includegraphics[width=8.8cm]{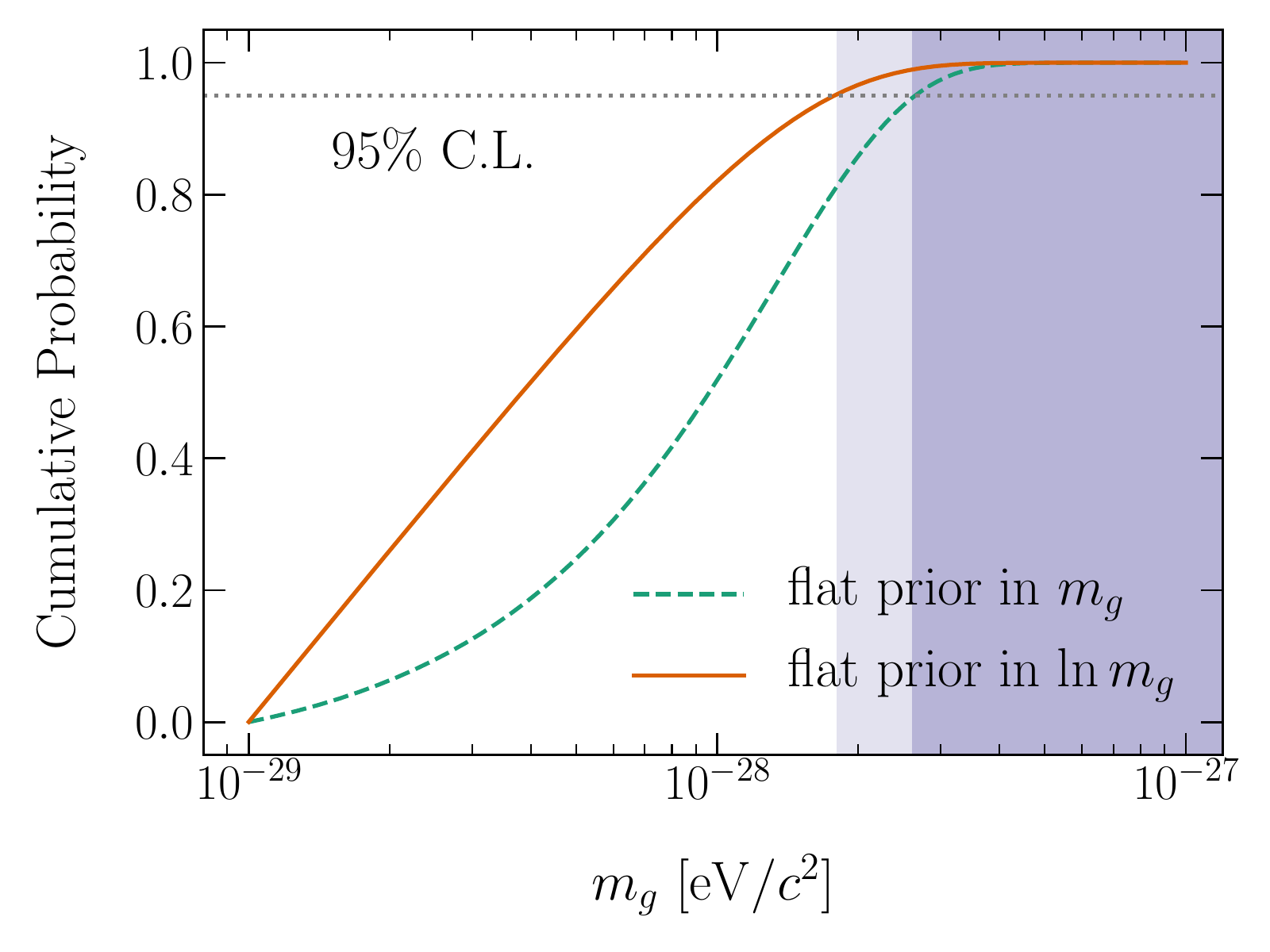}
  \caption{Cumulative probability for the graviton mass with flat priors in
  $m_g$ ({\it dashed}) and $\ln m_g$ ({\it solid}). Shaded regions show the
  excluded graviton mass values at 95\% C.L. [see Eq.~(\ref{eq:limit:mg})
  and Eq.~(\ref{eq:limit:lnmg})].}
  \label{fig:cdf}
\end{figure}

All the pulsars in the table have passed the tests of
GR~\cite{Wex:2014nva}. \ReplyA{In particular, the measured orbital decay
rates, after subtracting the kinematic Shklovskii effect and the Galactic
acceleration contribution, agree with the GR prediction in
Eq.~(\ref{eq:Pbdot:GR}) within uncertainty for all binary pulsars that we
consider.} Therefore, here we do not look for {\it evidence} of the
Galileon radiation, instead we put {\it upper bounds} on the graviton mass
via constraining the extra Galileon radiation. By plainly assuming that the
Galileon radiation is smaller than the uncertainty in $\dot P_b$, we obtain
upper bounds for the graviton mass for each pulsar. These bounds are
plotted in Fig.~\ref{fig:constraint} for cases using $\sigma_{\dot P_b^{\rm
int}}$ and $\sigma_{\dot P_b^{\rm obs}}$, as function of the orbital period
$P_b$ and the orbital eccentricity $e$.

The scenario of using $\sigma_{\dot P_b^{\rm int}}$ should be taken as
providing the current bounds on $m_g$. In this scenario, the best bound is
given by the agreement of $\dot P_b$ to the GR prediction within $10^{-3}$
from the recent observation of the double pulsar
PSR~J0737$-$3039A~\cite{Kramer:2016kwa}. The bound on $m_g$ reads,
\begin{equation}
  \ReplyB{m_g \lesssim 3 \times 10^{-28} \, {\rm eV}/c^2 \quad \mbox{(95\% C.L.)} } \,,
\end{equation}
which is two times better than the bound from the Hulse-Taylor pulsar
PSR~B1913+16, which gives \ReplyB{$m_g \lesssim 6\times10^{-28}\, {\rm eV}/c^2$
(95\% C.L.)~\cite{Weisberg:2016jye}.} Although the larger eccentricity of
PSR~B1913+16 ($e \simeq 0.617$) is beneficial to the test, compared with a
mildly small eccentricity for PSR~J0737$-$3039A ($e \simeq 0.088$), the
astrophysical contribution introduces a significant uncertainty to the
intrinsic $\dot P_b$ of PSR~B1913+16~\cite{Damour:1990wz}. The double
pulsar, being relatively close to the Earth, is not yet limited by the
astrophysical contribution in $\dot P_b$~\cite{Kramer:2006nb}.

The scenario of using $\sigma_{\dot P_b^{\rm obs}}$, on the other hand,
should be taken as over-optimistic, representing the ability of {\it clean}
binary pulsar systems in constraining the graviton mass. This is only
possible when the kinematic Shklovskii effect is measured and the Galactic
acceleration is modeled both to be better than the observational
uncertainty. In such an idealized case, the best pulsar would be
PSR~B1534+12~\cite{Fonseca:2014qla}, which gives $m_g \lesssim 6 \times
10^{-29} \, {\rm eV}/c^2$ (95\% C.L.) by itself alone. These
over-optimistic results are also plotted in Fig.~\ref{fig:constraint} with
orange circles, but they are only considered to be indicative of technology
capability and not used in the following analysis.

Because the theory parameter, $m_g$, is universal to different pulsars, we
can combine the constraints from individual pulsars into a combined
constraint. We assume that the measurements for different binary pulsars
are independent, thus the covariant matrix is diagonal for them. \ReplyB{We
make use of a simple (logarithmic) likelihood $ \ln {\cal L} \equiv -
\frac{1}{2} \sum_i \big( \dot P_b^\pi / \sigma_{\dot P_b^{\rm int}} \big)^2
$, where $\dot P_b^\pi$ is a sum of Galileon radiations in
Eqs.~(\ref{eq:Pbdot:mono}--\ref{eq:Pbdot:quad}), $\sigma_{\dot P_b^{\rm
int}}$ is given in Table~\ref{tab:psr}, and the summation is over all
binary pulsars considered in this paper.} We can also assign prior
knowledge to the graviton mass within the framework of Bayesian
statistics~\cite{DelPozzo:2016ugt}. In Fig.~\ref{fig:cdf} we plot the
combined cumulative probability distribution of the graviton mass for two
different sets of prior knowledge. \ReplyB{The combined bound is dominantly
influenced by PSRs~J0737$-$3039A and B1913+16, while the other twelve
pulsars only play a minor role.} From the figure, we obtain
\begin{equation}\label{eq:limit:mg}
  \ReplyB{ m_g \lesssim 3 \times 10^{-28} \, {\rm eV}/c^2 \quad \mbox{(95\% C.L.)}} \,,
\end{equation}
when a uniform prior on $m_g \in ( 10^{-29}, 10^{-27} ) \, {\rm
eV}/c^2 $ is assumed, and
\begin{equation}\label{eq:limit:lnmg}
  \ReplyB{m_g \lesssim 2 \times 10^{-28} \,  {\rm eV}/c^2 \quad \mbox{(95\% C.L.)}} \,,
\end{equation}
when a uniform prior on $\ln m_g$ for $m_g \in ( 10^{-29}, 10^{-27} ) \,
{\rm eV}/c^2 $ is assumed; the high end of the prior range, namely
$10^{-27}\, {\rm eV}/c^2$, comes from the analysis in
Ref.~\cite{deRham:2012fw}. While the former bound (\ref{eq:limit:mg}) is
very robust, the latter bound~(\ref{eq:limit:lnmg}) using a uniform prior
on $\ln m_g$ depends on our choice of prior range. It is a generic feature
of Bayesian analysis. In a cosmologically favored reasoning, one might
expect the graviton mass to be around the current Hubble scale, namely $m_g
\sim H_0 \sim 10^{-33}\,{\rm eV}/c^2$. If we had used a uniform prior on
$\ln m_g$ for $m_g \in ( 10^{-34}, 10^{-27} ) \, {\rm eV}/c^2$, we would
have obtained a much tighter bound of $m_g$. However, such a bound is
dominated by the prior knowledge. It means that binary pulsars are not yet
sensitive to a cosmologically small graviton mass. Therefore, we stick to
our {\it relatively} conservative result in Eq.~(\ref{eq:limit:lnmg}). In
contrast, the above change in the prior range does not affect the bound in
Eq.~(\ref{eq:limit:mg}) when a uniform prior on $m_g$ is adopted.

\ReplyB{The results on $m_g$ here improve the one in
Ref.~\cite{deRham:2012fw}, $m_g \lesssim 10^{-27}\,{\rm eV}/c^2$ (95\%
C.L.)}, due to the use of an updated analysis and the use of recent
observational results of pulsar timing, in particular the use of the double
pulsar results given in Ref.~\cite{Kramer:2016kwa}.

\section{Projected constraints with pulsar-BH systems}
\label{sec:bhs}

A well-timed pulsar around a BH companion is a long-sought holy grail in
the pulsar astronomy. The discovery of these systems will enable a couple
of unprecedented tests of gravity theories, in particular on the aspects
related to the property of BH solutions~\cite{Wex:1998wt, Kramer:2004hd,
Shao:2014wja, Shao:2018klg, Seymour:2018bce, Bull:2018lat}. Up to now,
despite extensive dedicated searches, no conclusively convincing candidate
has been found. The uncertainty in the estimation of the number of these
potential systems pertains mainly to their formation channels. We will not
discuss further the involved astrophysics here. Nevertheless, several
studies have shown a good potential for the discovery of pulsar-BH systems
in the near future~\cite{Liu:2011ae, Wharton:2011dv, Liu:2014uka,
Bower:2018mta, Bower:2019ads}. If such a pulsar-BH system is discovered, it
will provide a completely new playground to perform interesting tests of
gravity, including the Galileon radiation~\cite{Seymour:2018bce}.

\begin{figure}[t]
  \includegraphics[width=8.8cm]{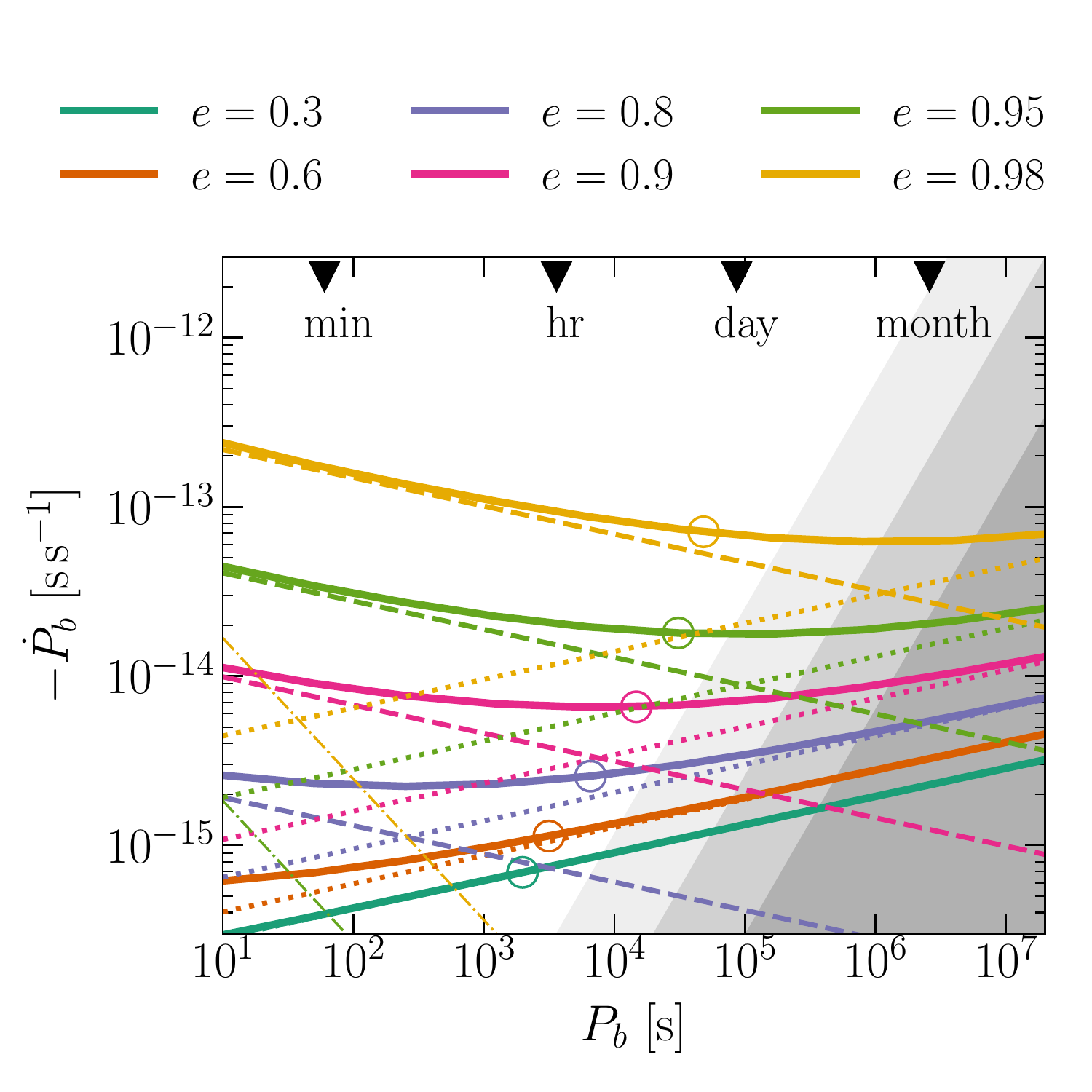}
  \caption{Same as Fig.~\ref{fig:Pb}, but for $m_1=1.4\,M_\odot$,
  $m_2=10\,M_\odot$, and $m_g = 10^{-28}\,{\rm eV}/c^2$. Open circles
  represent systems that have lifetime of 1\,Myr before merger in GR.
  Shaded regions show the estimated measurement precision that can be
  achieved with (from dark to light) $\sigma_{\rm TOA} = 0.1\,\mu$s,
  $1\,\mu$s, and $10\,\mu$s; the precision is found to be independent of
  eccentricity. Note that our simulations are performed for orbits
  with $P_b \gtrsim 0.2\,$day, and in this figure orbits with $P_b \lesssim
  0.2\,$day are extrapolated with Eq.~(\ref{eq:simulation:fit}).} \label{fig:BHs}
\end{figure}

In Fig.~\ref{fig:BHs}, we plot the expected contribution from the Galileon
radiation to the orbital decay rate for a pulsar -- $10\,M_\odot$ BH
system. In this figure, the graviton mass is assumed to be $m_g =
10^{-28}\,{\rm eV}/c^2$, which is roughly the best bound from the
combination of the current binary pulsars, obtained in
Eqs.~(\ref{eq:limit:mg}) and (\ref{eq:limit:lnmg}). Because of the
asymmetric masses, for extreme (unrealistic) systems with $P_b \lesssim
1\,{\rm minute}$ and $e \gtrsim{0.95}$, the Galileon dipole radiation could
be relevant to the total radiation. However, even if there exist such kind
of systems, they are unlikely to be detected with radio telescopes due to
the large orbital acceleration and limited computation
resources~\cite{Liu:2014uka}; however, fast imaging and imaging searches
based on significant circular polarization or scintillation might
help~\cite{Law:2018, Kaplan:2019psa, Dai:2016}. The expected number for the
LISA detector is very low as well, due to their short lifetime before the
merger~\cite{Audley:2017drz}. For binaries with $P_b \gtrsim 1\,{\rm day}$,
the Galileon quadrupole radiation is still the dominant contributor.

Depending on the eccentricity, now the Galileon monopole radiation could
play an essential role for binaries with $P_b \lesssim 1\,{\rm day}$.
Discovery of such binaries might be realistic for current and upcoming
radio telescopes~\cite{Kramer:2004hd, Liu:2014uka}. \citet{Liu:2014uka}
conducted extensive time-of-arrival (TOA) simulation for a pulsar-BH system
with masses $\left(m_1, m_2\right) = \left(1.4, 10\right) \, M_\odot$.

Similar to Ref.~\cite{Liu:2014uka}, we have conducted extensive mock-data
simulations, for different pulsar-BH configurations. For all these
simulations we have assumed one observing session per week with ten TOAs of
given uncertainty $\sigma_{\rm TOA}$, over a period of five years. \ReplyB{We further assume that the TOAs follow a Gaussian distribution and are uncorrelated (white noise).} Our simulations cover an orbital period range from 0.2 to about 100 days. We find
that, with given orbital period, the precision in measuring $\dot P_b$,
denoted by the (dimensionless) quantity $\sigma_{\dot P_b}$, is only weakly
dependent on the orbital eccentricity. The dependence on $P_b$ is nicely
fitted by,
\begin{equation} \label{eq:simulation:fit}
  \log_{10} \sigma_{\dot P_b}= {\cal A} + {\cal B} \log_{10} \left(
  \frac{P_b}{\rm day} \right) \,,
\end{equation}
where $({\cal A}, {\cal B}) = (-15.67, 1.331)$, $(-14.67, 1.331)$, and
$(-13.68, 1.332)$, when $\sigma_{\rm TOA}$ is $0.1\,\mu$s, $1\,\mu$s, and
$10\,\mu$s, respectively. In fact, to good approximation one can use
${\cal A} = -14.7 + \log_{10}\left(\sigma_{\rm TOA} / \mu{\rm
s}\right)$ and ${\cal B} = 1.33$. \ReplyB{The actual uncertainty obtained for a TOA depends on various aspects, like pulsar luminosity, pulse profile, integration time, telescope and backend parameters, etc. The assumed uncertainties and number of TOAs are typical for precision timing observations in pulsar astronomy (see e.g.\ Ref.~\cite{Perera:2019sca}).} The expected precision for the three
different values of $\sigma_{\rm TOA}$ is shown as shaded regions in
Fig.~\ref{fig:BHs}. We see that pulsar-BH systems will have a great
potential to improve the current best bound on $m_g$. For example, if
$\sigma_{\rm TOA}$ about $0.1\,\mu$s ($10\,\mu$s) is achieved, an eccentric
pulsar with $P_b \lesssim 1$\,month ($P_b \lesssim 1$\,day) can probe $m_g$
down to the level of $10^{-28}\,{\rm eV}/c^2$. We would like to emphasise
that the timing precision assumed here can generally only be achieved for
recycled pulsars, even with large radio telescopes like the
Five-hundred-meter Aperture Spherical Telescope (FAST)~\cite{Jiang:2019,
Lu:2019} or the upcoming Square
Kilometre Array (SKA)~\cite{Kramer:2004hd, Shao:2014wja,Bull:2018lat}. Such
pulsar-BH systems could in principle be the result of an exchange encounter
in regions with high stellar density, like globular clusters and the
Galactic centre region (see e.g.~Ref.~\cite{fl11}). The actual timing
precision will, in addition, depend on other parameters like pulsar
luminosity, pulse shape, etc.

\begin{figure}[t]
  \includegraphics[width=8.8cm]{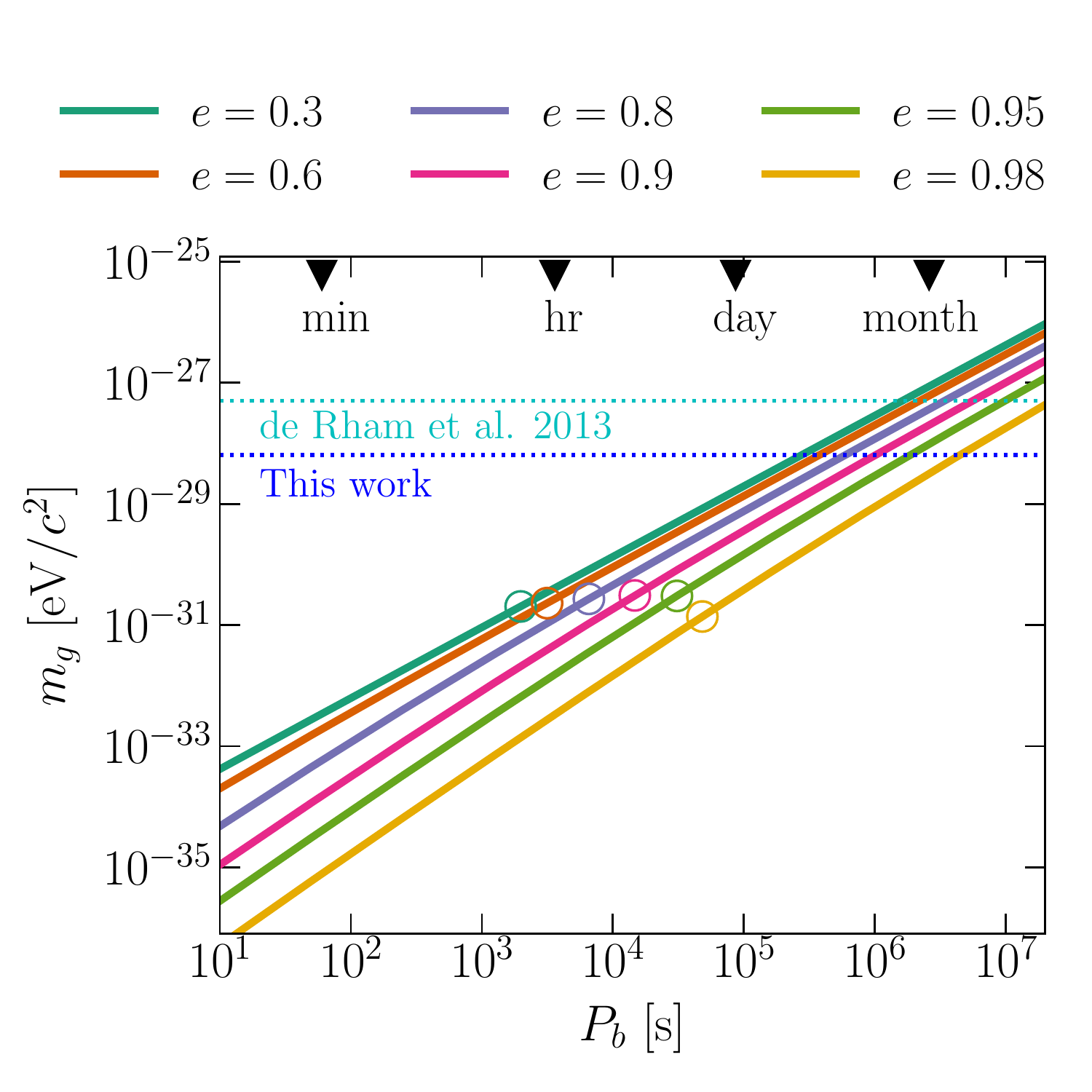} 
  \caption{Projected bounds \ReplyB{at 68\% C.L.} on $m_g$ from a pulsar-BH
  system with $\sigma_{\rm TOA} = 0.1\,\mu$s. Open circles represent
  systems that have lifetime of 1\,Myr before merger in GR. As in
  Fig.~\ref{fig:BHs}, results for orbits with $P_b \lesssim 0.2\,$day are
  extrapolated via Eq.~(\ref{eq:simulation:fit}).} \label{fig:SKA}
\end{figure}

In Fig.~\ref{fig:SKA}, we plot the projected bounds on $m_g$ using the
$\dot P_b$ precision in Eq.~(\ref{eq:simulation:fit}) with $\sigma_{\rm
TOA}=0.1\,\mu$s. As we can see from the figure, if we can time a
near-circular pulsar-BH binary with $P_b$ smaller than a few days, the
bound (\ref{eq:limit:lnmg}) in this paper can be improved. If the binary is
highly eccentric, then an orbital period $P_b$ smaller than a few months is
sufficient to improve the bound, as was indicated in Fig.~\ref{fig:BHs}. On
the other hand, if one wants to probe the cosmologically interested range
for $m_g \sim 10^{-33}\,{\rm eV}/c^2$~\cite{deRham:2012fw}, a sub-minute
circular binary, or a sub-hour highly eccentric binary, is needed. We
consider such cases highly unlikely to be discovered with near-future
technologies, leaving aside the fact that the existence of such a system in
our Galaxy is almost certainly excluded, due to its short merger time.

\begin{figure}[t]
  \includegraphics[width=8.8cm]{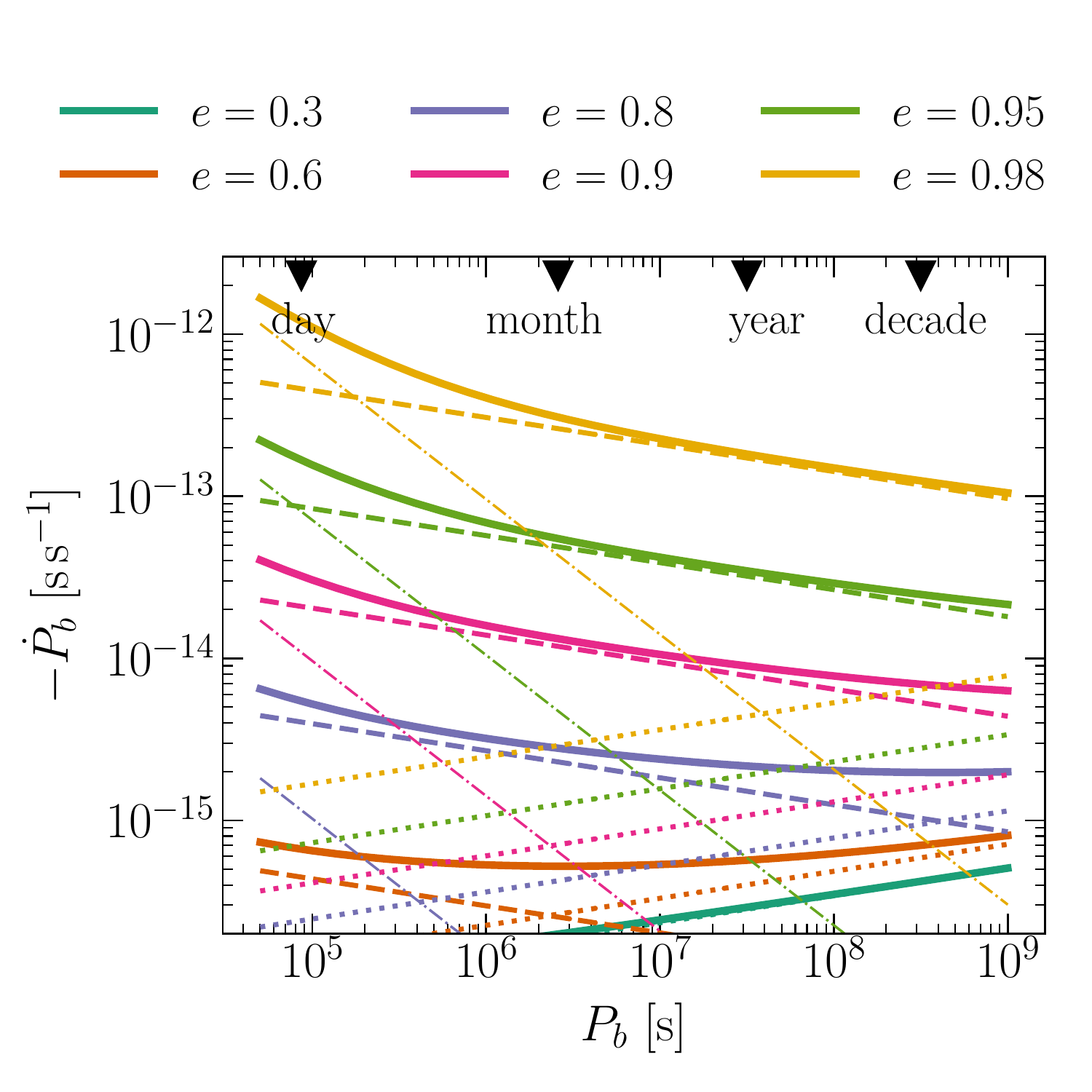}
  \caption{Same as Fig.~\ref{fig:Pb}, but for $m_1=1.4\,M_\odot$,
  $m_2=4\times10^6\,M_\odot$, and $m_g = 10^{-28}\,{\rm eV}/c^2$.}
  \label{fig:SgrA}
\end{figure}

In another direction, extensive searches for pulsars around the Sgr\,A$^*$,
the supermassive BH at the centre of our Galaxy, are
ongoing~\cite{Liu:2011ae, Bower:2018mta, Bower:2019ads, Goddi:2017pfy}. In
Fig.~\ref{fig:SgrA} we plot the Galileon radiation for a pulsar --
Sgr\,A$^*$ BH system for different eccentricities with $m_g =
10^{-28}\,{\rm eV}/c^2$. For eccentric systems with $P_b \lesssim 1\,{\rm
yr}$, the Galileon monopole radiation prevails over the Galileon quadrupole
radiation for $e \gtrsim{0.6}$. Therefore, there might be an opportunity to
test the Galileon monopole radiation within this class of systems. To
address this question, we perform mock-data simulations to investigate the
precision one can expect for $\dot P_b$ for a pulsar in a suitable orbit
around Sgr A*. Similar to Ref.~\cite{Liu:2011ae}, we have created mock data
with one TOA of 100\,$\mu$s precision every week, over a time span of five
years. Furthermore, we have assumed that the pulsar orbit is unperturbed
and therefore our parameter estimation is based on a phase-connected timing
solution that provides a perfect fit over the whole time span of
observations. Even under such optimistic assumptions, we find that, for an
orbit with $P_b = 0.5\,$yr and $e=0.8$, it is unlikely to get a
$\sigma_{\dot P_b}$ better than $10^{-12} \,{\rm s\,s}^{-1}$. Therefore,
only in the event of a pulsar in a highly eccentric orbit with $P_b \lesssim 1\,$day being
discovered, we might be able to improve the bound (\ref{eq:limit:lnmg}).
However, we consider the existence of a pulsar in such an orbit extremely
unlikely.

The tests in this section depend sensitively on the actual eccentricity of
the pulsar-BH system, as shown in the Figs.~\ref{fig:BHs}--\ref{fig:SgrA}.
In fact, the analysis with Fig.~\ref{fig:BHs} and Fig.~\ref{fig:SgrA} is
conservative, because the Galileon radiation power was calculated averaging
over the orbital timescale~\cite{deRham:2012fw}. In reality, during the
periastron passage the gravitational radiation will be maximized, and
produce prominent features. These features are believed to provide even
better distinguishable signals. An analysis resolving the orbital timescale
is beyond the scope of this paper. Because all three kinds of the Galileon
radiations are proportional to $m_g$, the results discussed in
Fig.~\ref{fig:BHs} and~\ref{fig:SgrA} can be rescaled easily with different
graviton mass for different binary systems.

\section{Discussion}\label{sec:disc}

In this paper we systematically studied the Galileon radiation in cubic
Galileon theory~\cite{deRham:2012fw} in the context of pulsar timing.
Because the Galileon radiation is screened differently than its fifth-force
counterpart, such a study is essential to better understand the basic
role of the Vainshtein mechanism in screening the gravitational
radiation. 

From a collection of fourteen well-timed binary pulsars, we have obtained a
new bound on the theory parameter for cubic Galileon, namely the graviton
mass,
\begin{equation}\label{eq:limit}
  \ReplyB{m_g \lesssim 2 \times 10^{-28} \, {\rm eV}/c^2 \quad \mbox{(95\% C.L.)} } \,,
\end{equation}
when a uniform prior on $\ln m_g $ for $m_{g} \in(10^{-29}, 10^{-27})
\,\mathrm{eV} / c^{2}$ is used. \ReplyB{This improves a previous bound from the
Hulse-Taylor pulsar~\cite{deRham:2012fw} by a factor of five.} Though the
bound (\ref{eq:limit}) is weaker than a few other bounds such as bounds
from the Earth-Moon-Sun system and dark matter
clusters~\cite{deRham:2016nuf}, it is nevertheless a robust bound from a
completely different regime, namely from the dynamic gravitational
radiation instead of the static environments. It is also immune from
uncertain assumptions about the dark matter distributions and the
virialization of the gravitating systems. Therefore, we consider this bound
{\it complementary} to the existing bounds.

Finally, \citet{deRham:2012fg} have discussed radiations from a generic
Galileon theory with all allowed interactions in the 4-dimensional
spacetime. The inclusion of quartic or quintic Galileon complicates the
calculations considerably. The authors found that, the naive perturbation
theory predicts divergent results in the radiation power when these
higher-order terms are considered, meaning that the perturbations
themselves are nonlinear. Partial results were obtained for binary systems
with specific assumptions about the screening lengthscales. In particular,
only circular binary orbits were analyzed, thus not applicable to most of
the systems that we consider in this paper. \ReplyB{For circular orbits
that were considered in Ref.~\cite{deRham:2012fg}, meaningful bounds can be
derived when there is a hierarchy of strong coupling scales.} We wish to
perform a more complete analysis with higher-order Galileon interactions in
a future study.

\acknowledgments

We are grateful to Paulo Freire, Xueli Miao, Robert Wharton, and Kent Yagi
for helpful discussions, and Vivek Venkatraman Krishnan for carefully
reading the manuscript. This work was supported by the National Natural
Science Foundation of China (11975027, 11991053, 11721303), the Young Elite
Scientists Sponsorship Program by the China Association for Science and
Technology (2018QNRC001), the Max Planck Partner Group Program funded by
the Max Planck Society, and the High-performance Computing Platform of
Peking University. LS and NW acknowledge support from the European Research
Council (ERC) for the ERC Synergy Grant BlackHoleCam under Contract
No.~610058. SYZ acknowledges support from the starting grants from
University of Science and Technology of China under grant No.~KY2030000089
and GG2030040375 and is also supported by National Natural Science
Foundation of China (NSFC) under grant No.~11947301.

\bibliography{refs}

\end{document}